\documentclass[
 reprint,
 amsmath,amssymb,
 nofootinbib,
 aps,
 prd,
 superscriptaddress,
]{revtex4-1}
\usepackage{booktabs}
\usepackage{multirow}
\usepackage{aas_macros}
\usepackage{dcolumn}
\usepackage{bm}
\usepackage{amsmath}
\usepackage{float}
\usepackage{lipsum}
\usepackage{color}
\usepackage{times}
\usepackage{textcomp}
\usepackage{latexsym}
\usepackage[hidelinks]{hyperref}
\usepackage{acro}
\usepackage{mathtools}
\usepackage{url}
\usepackage[utf8]{inputenc}
\usepackage{physics}
\usepackage[dvipsnames]{xcolor}
\usepackage{soul}
\usepackage[caption = false]{subfig}
\usepackage[]{graphicx}
\usepackage{multirow}
\usepackage{booktabs}
\usepackage{adjustbox}
\usepackage[normalem]{ulem}

\usepackage [english]{babel}
\usepackage [autostyle, english = american]{csquotes}
\MakeOuterQuote{"}

\newcommand{\approptoinn}[2]{\mathrel{\vcenter{
  \offinterlineskip\halign{\hfil$##$\cr
    #1\propto\cr\noalign{\kern2pt}#1\sim\cr\noalign{\kern-2pt}}}}}

\newcommand{\FAP}{\text{FAP}}
\newcommand{\FAPpair}{$\rm FAP_{per\ pair}\ $}
\newcommand{\FAPtriple}{$\rm FAP_{per\ triplet}$}
\newcommand{\FAPquadruple}{$\rm FAP_{per\ quadruplet}$}

\begin{document}

\title{Lensing or luck? False alarm probabilities for gravitational lensing of gravitational waves}
\date{\today} 

\author{Mesut \c{C}al{\i}\c{s}kan}
\email{caliskan@jhu.edu}
\affiliation{The William H. Miller III Department of Physics and Astronomy, Johns Hopkins University, Baltimore, MD 21218, USA}
\affiliation{Kavli Institute for Cosmological Physics and Enrico Fermi Institute, The University of Chicago, Chicago, IL 60637, USA}

\author{Jose Mar\'{i}a Ezquiaga}
\email{ezquiaga@uchicago.edu, NASA Einstein fellow}
\affiliation{Kavli Institute for Cosmological Physics and Enrico Fermi Institute, The University of Chicago, Chicago, IL 60637, USA}
\affiliation{Niels Bohr International Academy, Niels Bohr Institute, Blegdamsvej 17, DK-2100 Copenhagen, Denmark}

\author{Otto A. Hannuksela}
\email{oahannuksela@cuhk.edu.hk}
\affiliation{Department of Physics, The Chinese University of Hong Kong, Shatin, NT, Hong Kong}

\author{Daniel E. Holz}
%\email{holz@uchicago.edu}
\affiliation{Kavli Institute for Cosmological Physics and Enrico Fermi Institute, The University of Chicago, Chicago, IL 60637, USA}
\affiliation{Department of Physics, and Department of Astronomy and Astrophysics, The University of Chicago, Chicago, IL 60637, USA}

\begin{abstract}
\noindent
Strong gravitational lensing of gravitational waves (GWs) has been forecasted to become detectable in the upcoming LIGO/Virgo/KAGRA observing runs.
However, definitively distinguishing pairs of lensed sources from random associations is a challenging problem.
We investigate the degree to which unlensed events mimic lensed ones because of the overlap of parameters due to a combination of random coincidence and errors in parameter estimation.  Lensed events are expected to have consistent masses and sky locations, and constrained relative phases, but may have differing apparent distances due to lensing magnification.
We construct a mock catalog of lensed and unlensed events.
We find that the probability of a false alarm based on coincidental overlaps of the chirp mass, sky location, and coalescence phase are approximately $9\%$, $1\%$, and $10\%$ per pair, respectively.
Combining all three parameters, we arrive at an overall false alarm probability per pair of $\sim10^{-4}$.
We validate our results against the GWTC-2 data, finding that the catalog data is consistent with our simulations.
As the number of events, $N$, in the GW catalogs increases, the number of random pairs of events increases as $\sim N^2$. 
Meanwhile, the number of lensed events will increase linearly with $N$, implying that for sufficiently high $N$, the false alarms will always dominate over the true lensing events. This issue can be compensated for by placing higher thresholds on the lensing candidates (e.g., selecting a higher signal-to-noise ratio (SNR) threshold, $\rho_{\rm thr}$), which will lead to better parameter estimation and, thus, lower false alarm probabilities per pair---at the cost of dramatically decreasing the size of the lensing sample ($\propto \rho_{\rm thr}^{-3}$). We show that with our simple overlap criteria for current detectors at design sensitivity, the false alarms will dominate for realistic lensing rates ($\lesssim10^{-3}$) even when selecting the highest SNR pairs. 
These results highlight the necessity to design alternative identification criteria beyond simple waveform and sky location overlap for conclusive detection of strong lensing.
Future GW detectors such as Cosmic Explorer and Einstein Telescope may provide sufficient improvement in parameter estimation and a commensurate decrease in the incidence of coincidental overlap of parameters, allowing for the conclusive detection of strong lensing of GWs even without additional detection criteria.
\end{abstract}

\maketitle

\section{Introduction}

\noindent
When light travels near massive objects over cosmological distances, it is gravitationally lensed~\cite{1998PhRvD..58f3501H,Bartelmann:2010fz}.
This phenomenon leads to many interesting observations in the electromagnetic band, such as distortions of galaxy images into long arcs or Einstein "rings," time-variable changes in the flux emitted by stars in the limit of micro-lensing, multiple images of the same supernova explosion, and statistical distortions of background light in the limit of weak lensing.
Gravitational lensing of electromagnetic waves is widely used in studies of astronomy, astrophysics, and cosmology, including constraining the dark matter~\citep{Clowe:2003tk,Markevitch:2003at}, discovering exoplanets~\citep{Bond:2004qd}, measuring the Hubble constant~\citep{2016A&ARv..24...11T}, uncovering massive objects and structures that are too faint to be detected directly~\citep{1999A&A...351L..10S,Coe:2012wj}, and a wide range of other effects (e.g., see~\citep{1998ApJ...506L...1H,1999ApJ...510...54H,2001ApJ...556L..71H,2003ApJ...585L..11D,2006PhRvL..96b1301C,2007PhRvD..76l7301H}). 

Just as in the case of light, gravitational waves (GWs) can also be gravitationally lensed~\cite{Ohanian:1974ys,Thorne:1982cv,Deguchi:1986zz,Wang:1996as,Nakamura:1997sw,Takahashi:2003ix}. 
However, the tools to detect and analyze lensed GWs have been developed only recently~\cite{Cao:2014oaa,Lai:2018rto,Christian:2018vsi,Hannuksela:2019kle,Li:2019osa,McIsaac:2019use,Pang:2020qow,Hannuksela:2020xor,Dai:2020tpj,Liu:2020par,Wang:2021kzt,Lo:2021nae,Janquart:2021nus,Janquart:2021qov}. 
In particular, when a GW is strongly lensed, its amplitude, arrival time, and overall phase can change while the frequency evolution remains the same~\citep{Ohanian:1974ys,Thorne:1982cv,Deguchi:1986zz,Wang:1996as,Nakamura:1997sw,Takahashi:2003ix,Dai:2017huk,Ezquiaga:2020gdt}.
As a consequence, GW detectors might detect multiple images as repeated events separated by time delays of minutes to months when lensed by galaxies~\citep{Ng:2017yiu,Li:2018prc,Oguri:2018muv} and up to years when lensed by galaxy clusters~\citep{Smith:2017mqu,Smith:2018gle,Smith:2019dis,Robertson:2020mfh,Ryczanowski:2020mlt}.
If the GWs propagate near smaller lenses such as stars or compact objects, micro- or milli-lensing may produce observable frequency-dependent amplitude and phase modulations~\citep{Cao:2014oaa,Jung:2017flg,Lai:2018rto,Christian:2018vsi, Diego:2019lcd,Diego:2019rzc,Pagano:2020rwj,Caliskan:2022hbu}.
Indeed, lensing can induce many effects on GWs.

If observed, lensed GWs would enable a plethora of new scientific studies. 
For example, they might allow us to locate merging black holes at sub-arcsecond precision when combined with electromagnetic lensing surveys~\cite{Hannuksela:2020xor}.
When accompanied by an electromagnetic counterpart, they could enable precision cosmography studies owing to the sub-millisecond lensing time-delay measurements granted by GW observations~\cite{Sereno:2011ty, Liao:2017ioi, Cao:2019kgn, Li:2019rns, Hannuksela:2020xor, Yu:2020agu}.
Lensing of GWs could be used to perform precision tests of the speed and polarization content of GWs~\cite{Baker:2016reh, Fan:2016swi,Goyal:2020bkm} and to break the so-called mass-sheet degeneracy~\cite{Cremonese:2021puh}. 
They could also be used to detect intermediate-mass and primordial black holes through micro-lensing observations~\cite{Lai:2018rto, Diego:2019rzc, Oguri:2020ldf}. Finally, the statistics of lensing distributions may provide a clean and powerful cosmological probe, constraining both the source and lens populations~\cite{Xu:2021bfn}.

\begin{figure}[t!]
    \centering
    \includegraphics[width=\columnwidth]{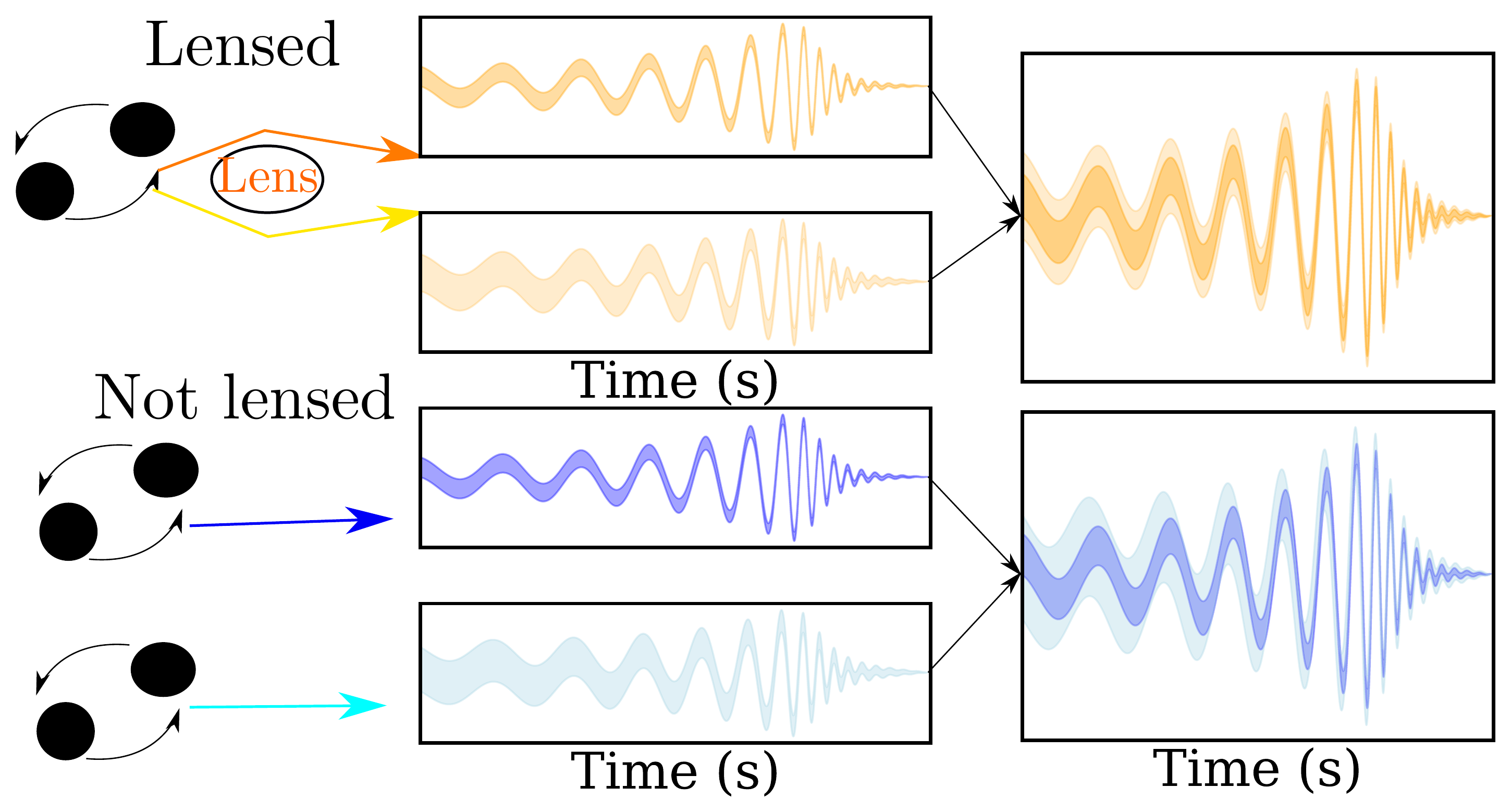}
    \caption{Comparison of waveforms of lensed images from the same event (upper panels) versus waveforms of images from separate events (lower panels). The waveforms of the lensed images show high resemblance, although their amplitudes are different due to the magnification caused by lensing. On the other hand, the waveforms of the unlensed images also show resemblance (due to coincidence), mimicking lensing.
    }
    \label{fig:mimickedLensing}
\end{figure}

A particularly exciting area is the strong lensing of GWs, forecasted to be at a detectable rate at design sensitivity~\cite{Ng:2017yiu,Li:2018prc,Oguri:2018muv, Xu:2021bfn, Mukherjee:2021qam, Wierda:2021upe}. 
Multiply imaged GW sources produce two or more waveforms with consistent frequency evolution and sky locations but different overall amplitudes, arrival times, and complex phases~\cite{Haris:2018vmn}. 
To detect such near-identical signal pairs, we test if two waveforms have identical detector-frame parameters within detector accuracy~\cite{Haris:2018vmn,Hannuksela:2019kle,Dai:2020tpj,Liu:2020par,Lo:2021nae}. 
However, it is also possible for two waveforms to have similar detector-frame parameters by coincidence, resulting in a "false alarm."
Figure~\ref{fig:mimickedLensing} shows an example in which two unlensed waveforms from separate events show high resemblance due to coincidence, mimicking lensing.
This is a timely problem to investigate, given that several groups have reported the intriguing possibility that LIGO~\cite{LIGO_2015} and Virgo~\cite{Virgo_2014} have already observed lensing~\cite{Broadhurst:2018saj,Broadhurst:2019ijv,Broadhurst:2020moy,Dai:2020tpj}, and in light of the recent LIGO--Virgo--Kagra search for lensing~\cite{LIGOScientific:2021izm}.

To show that an event pair is lensed, we must rule out the possibility that it is just a chance overlap instead of multiple images of a single source. 
Two approaches exist to distinguish these possibilities. 
First, statistical approaches rely on quantifying the likelihood of a false alarm using a particular metric, usually called coherence ratio, and a simulated population of events~\cite{Haris:2018vmn,Dai:2020tpj,Liu:2020par,Janquart:2021qov}.
A second approach quantifies the likelihood that a given event pair is lensed using a Bayes' factor, which relies on accurate modeling of the source and lens populations~\cite{Lo:2021nae}. 
However, prior literature has not focused on (i) understanding to what extent individual parameters contribute to the false alarm rate, (ii) quantifying which types of events would enable more confident detection of lensing, and (iii) how triply and quadruply-imaged events raise our confidence in lensed detections. 
In what follows we address these issues. 

In particular, we investigate the probability that random astrophysical processes could create two GWs sources that mimic strong lensing. 
We construct a mock catalog of lensed and unlensed events, and define our parameter overlap statistics, in Sec.~\ref{sec:methods}. 
We present the results for the overlap in our mock catalog for the mass, sky position, and phase in Sec.~\ref{sec:results}. 
We also discuss the total false alarm probability, effects of triple and quadruple images and higher signal-to-noise ratio (SNR) thresholds on the false alarm probability, and the necessary conditions for conclusive identification of strong lensing. 
We conclude in Sec.~\ref{sec:conclusions}.

%--------
%SECTION: METHODS
%--------
\section{Methods} \label{sec:methods}

%Intro lensing
The propagation of a wave around a lens can be obtained by solving the diffraction integral, $F$, which considers the propagation along all possible paths (e.g., see~\cite{Schneider:1992}). 
In the stationary phase approximation, when the time delay between the paths $\Delta t$ is larger than the inverse frequency of the wave, i.e., $|\omega\Delta t|\gg1$, this integral is dominated by the stationary points. This implies that distinct images are formed.
In this regime, for a frequency domain waveform $\tilde{h}(\omega)$, the lensed signal will be given by
\begin{align}\label{Fgeom}
    &\tilde{h}_L(\omega) = F(\omega) \tilde{h}(\omega)\,, \\
    &F(\omega)\approx \sum_j\vert\mu_j\vert^{1/2}\exp\left(i\omega \Delta t_j-i\, \text{sign}(\omega)\frac{n_j\pi}{2}\right),
\end{align}
where $\mu_j$ is the amplification factor of the $j$-th image, $\Delta t_j$ its time delay, and $n_j$ the Morse index. This last frequency-independent phase shift is associated with the image type or number of caustics crossed between the source and the observer. For a single lens it takes three possible values: $n_j=0$ for Type I, $n_j=1$ for type II, and $n_j=2$ for type III images. 
Therefore, in the regime of multiple images (also known as strong lensing), a GW source will suffer three main effects: (i) an amplitude change via $\mu_j$, (ii) a different arrival time due to $\Delta t_j$, and (iii) a possible waveform distortion due to $n_j$ if the original signal contains multiple frequency components. 
The first two modifications are exactly degenerate with a change in the luminosity distance and coalescence time. 
The last one is approximately degenerate with a change in the coalescence phase when the quadrupole radiation dominates the signal \cite{Dai:2017huk,Ezquiaga:2020gdt}. 

%Intro PE and search for overlaps
The fact that for ``typical'' binary black-hole mergers (those without significant higher modes, precession, or eccentricity) the effects of strong lensing are (approximately) degenerate with a few parameters describing the binary implies that we can search for multiple images by looking for correlations in the other parameters.\footnote{However, higher-order-modes can produce non-trivial waveform effects~\citep{Dai:2017huk,Ezquiaga:2020gdt, Lo:2021nae, Janquart:2021nus}; a joint analysis could reveal this~\cite{Lo:2021nae, Janquart:2021nus}.} 
In particular, one can search for events with the same masses, spins, and sky positions. 
Moreover, since the Morse index introduces only a fixed phase shift, one can also look for coalescence phase differences of multiples of $\pi/4$ ($\Delta\phi_\text{Morse}=2\Delta\varphi_c$). 

%Errors, selection biases and populations
The search for strongly lensed events is not as simple as this implies, however. The first obvious limitation is the observational error. 
Because some parameters are poorly measured (e.g., individual component spins), parameter overlap between unrelated events can happen due to astrophysical coincidence. 
A second difficulty is the impact of selection effects. 
Since ground-based detectors are more sensitive to certain frequencies and have a fixed geometry and location on Earth, they are preferentially sensitive to certain masses and sky positions, increasing the odds of chance overlap. 
Finally, there is uncertainty regarding the properties of the source population. If, for example, the binary population had a prominent peak at a certain mass, it would be expected to have many events overlapping precisely at this mass and thereby mimic lensing.

Given these characteristics of strong lensing and current GW detectors, in the following we assess how likely it is that an unlensed population has parameter overlaps by chance, and compare this rate to expectations for the true lensed population.
To achieve these goals, we first simulate a population of binary black holes (BBHs) as described in Sec.~\ref{sec:mock_catalog}. We then compute the overlap in their parameters using the metric presented in Sec.~\ref{sec:parameter_Overlaps}.
We calculate the false alarm probability based on the description presented in Sec.~\ref{sec:fap}, and discuss results in Sec.~\ref{sec:results}.

For simplicity, we consider the overlaps in mass, sky positions, and coalescence phase independently. 
Due to computational expense, a joint parameter estimation (PE) of the full parameter space of all possible event pairs (see~\citep{Liu:2020par,Lo:2021nae,Janquart:2021qov,Janquart:2021nus}) is not possible for the number of simulations we need to perform for this study. 
As we argue below, the number of potential lensing pairs increases as $\sim N^2$, and thus joint PE of all possible pairs rapidly becomes infeasible.
The false alarms quoted here should therefore be considered conservative, since including additional parameters could lower their significance.

%SUBSECTION: MOCK catalog
\subsection{Mock catalog of unlensed and lensed events} \label{sec:mock_catalog}

To simulate the population of sources, we consider simplified, parameterized models consistent with current observations. 
Our goal is to simulate posterior distributions for the masses, sky maps, and coalescence phase of a set of lensed and unlensed events. 

\paragraph{Simulating mass posteriors.} First, we generate a mock catalog of unlensed and lensed events to be used for mass overlaps.
We initially choose a simple model based on a power-law distribution and later investigate the effects of different populations on the lensing false alarm probability.
For the BBH component masses, we use a power-law mass function $p(m_1)\propto m_{1}^{-2.35}$ with $m_1 \in [5, 45]\  M_{\odot}$ for the primary mass, while the secondary mass $m_2$ is uniformly distributed in the mass ratio $q$, consistent with the black hole population properties inferred from LIGO/Virgo O1 and O2.~\citep{LIGOScientific:2018jsj}.
For the redshift evolution, we use a parameterized model 
\begin{equation}
    \mathcal{R}(z)=\mathcal{R}_0\, \mathcal{C}(\alpha,\beta,z_p)\frac{(1+z)^\alpha}{1+\left(\frac{1+z}{1+z_p}\right)^{\alpha+\beta}}
\end{equation}
designed to have a peak at $z_p = 1.9$, with a rising slope, $\alpha = 2.7$, and a decaying slope, $\beta = 2.9$. This choice corresponds to the BBHs following the star formation rate without a time delay~\cite{Madau:2014bja}.
Here, $\mathcal{R}_0$ is the local merger rate and $\mathcal{C}(\alpha,\beta,z_p)=1+(1+z_p)^{-\alpha-\beta}$,  
where we fix the background cosmology to {\em Planck}\/ 2018~\cite{Planck_2018}.  
We distribute the BBH mergers uniformly in the sky, and the inclination and polarization angles isotropically.
To compute the SNR, $\rho$, we use the phenomenological waveform model \texttt{IMRPhenomPv2}~\cite{Khan2016} and assume that the signals are detected by the LIGO--Virgo detector network at design sensitivity~\citep{aLIGOdesign, TheVirgo:2014hva}. 
We neglect the effect of spin in the SNR calculation. 

The population of strongly lensed mergers follows the same BBH distribution. 
However, we lens this BBH population presuming that the lenses are distributed as in the SDSS galaxy catalog and assuming the singular isothermal ellipsoid (SIE) model, following the approach in~\cite{Haris:2018vmn}. 
Concretely, this means that each binary (i) can be multiply imaged, (ii) has a probability $\tau(z_s)\propto V_c(z_s)$, where $V_{c}(z_s)$ is the comoving volume, of being lensed a priori (assuming that the density of lenses is constant in redshift), and (iii) can be magnified at magnification $\mu$, such that the lensed SNR becomes $\rho_L=\mu^{1/2} \rho$.

To assign measurement uncertainty to simulated observations described above, we use the following method based on~\citep{Fishbach_2020}. First, we determine the observed SNR, $\rho_{\rm obs}$, by
\begin{equation}
    \rho_{\rm obs} = \rho + n(0,1) \,,
\end{equation}
where $\rho$ is the actual SNR and $n(0,1)$ is a random number drawn from the standard normal distribution with mean 0 and standard deviation 1~\cite{maggiore2008gravitational}.
We assume that the sources are detected only if
\begin{equation}\label{eq:SNR_thres}
    \rho_{\rm obs} > \rho_{\rm thr} \,,
\end{equation}
where $\rho_{\rm obs}$ is the observed network SNR, and $\rho_{\rm thr}$ is the network SNR threshold. 
For a network of GW interferometers, the total network SNR $\rho_{\rm n}$ is defined as 
\begin{equation}
    \rho_\mathrm{n}^2=\sum_i\rho_i^2\,,
\end{equation}
where $\rho_{i}$ represents the SNR of the $i^{\rm th}$ GW interferometer. We calculate the observed network SNR based on the SNR of each GW interferometer as described above, and consider the network SNR threshold of $\rho_{\rm thr} = 12$.

We work both with
\begin{equation}
    \mathcal{M}_z=(1+z)\frac{(m_1m_2)^{3/5}}{(m_1+m_2)^{1/5}} \,,
\label{eq:ChirpMass}
\end{equation}
and
\begin{equation}
    M^{\rm total}_{z} = (1+z) (m_1 + m_2) \,,
\end{equation}
where $\mathcal{M}_z$ and $M^{\rm total}_{z}$ are the detector-frame (redshifted) chirp mass and detector-frame total mass, respectively. We assume that the uncertainties of the observed parameters scale inversely with $\rho_{\rm obs}$ so that
\begin{equation}
    \log(\mathcal{M}_z^{\rm obs}) = \log(\mathcal{M}_z) + n\left(0, \frac{\sigma_{\mathcal{M}}}{\rho_{\rm obs}} \right)\,,
\label{eq:mass_posterior}
\end{equation}
where $\mathcal{M}_z^{\rm obs}$ is the observed detector-frame chirp mass and $\sigma_{\mathcal{M}} = 0.053\rho_{\rm thr}$. We determine the posteriors of the detector-frame masses of the events in our mock catalogs based on Eq.~\ref{eq:mass_posterior}.

\paragraph{Simulating sky map posteriors.} We generate and localize a mock catalog of unlensed and lensed events to be used with sky map overlaps using \texttt{BAYESTAR}~\citep{PhysRevD.93.024013}, which is a Bayesian, non-Markov chain Monte Carlo sky localization tool that can rapidly localize a given GW event with accuracy similar to the parameter estimation of a full analysis.

For the unlensed population, we assume that the mergers have fixed masses with $m_1,\ m_2 = 30\,M_{\odot}$.\footnote{Fixing the masses for the simulations of the sky maps (as opposed to drawing them from a mass distribution as in the previous section) is a simplifying assumption to reduce the computational cost. It is justified by the fact that sky maps are dominated mainly by the SNR, and are relatively insensitive to the precise mass values. By choosing a representative SNR distribution, we can obtain a proper representation of the distribution of sky maps.} 
We also assume that the sky position and the inclination distributions are isotropic. 
We distribute the distances uniformly in volume in the range $0.07<z<1.9$~\citep{2021CQGra..38e5010C}.
We assume that the BHs have no spin. We generate the waveforms using \texttt{IMRPhenomPv2} and run the matched filter pipeline assuming the signals are detected by the LIGO-Virgo detector network at design sensitivity. 
We introduce Gaussian noise into the measurement. 
Similar to before, we choose the network SNR threshold to be $12$, and the minimum number of triggers to form a coincidence as $1$. 
We then localize the BBH mergers with \texttt{BAYESTAR} and form the sky maps.

For the lensed population, we make similar assumptions for the source parameters, except in this case we assume multiple images are detected for each merger.
We use this approach to estimate the degree to which lensed images of the same source overlap in their sky maps. This will be useful when we set a threshold overlap to cull out random overlaps for the unlensed population.
Like before, we impose consistent detection thresholds and sample the events until we have $\mathcal{O}(100)$ detected events. This approach is similar to those followed by, for example, Refs.~\cite{Ng:2017yiu, Wierda:2021upe, LIGOScientific:2021izm}. We localize the BBH mergers with \texttt{BAYESTAR} and form the sky maps.

\paragraph{Simulating coalescence phase posteriors.}
Lensed signals will differ in their overall phase. When restricted to typical signals, those dominated by the quadrupolar 22 mode, this corresponds to difference of an integer multiple of $\pi/4$ in the coalescence phase.
The coalescence phase itself is a quantity which is degenerate with other intrinsic parameters, and therefore the typical PE posterior for the phase is relatively unconstrained. 
However, when performing a joint PE analysis of a pair of events, many of these degeneracies are broken and one can constrain the phase difference.
For this reason, we focus directly on simulating the phase difference, $\Delta\varphi_c$. 
Performing a full joint PE analysis is beyond our simple analysis, and would not be expected to qualitatively change our results if we can anticipate the error in the phase difference.
As a benchmark, we benefit from the previous work of \cite{Dai:2020tpj} to estimate the typical error in the phase difference, and assume that the
$\Delta\varphi_c$ posterior is Gaussian. 

Once we have a model for the phase difference posterior, a population of unlensed events will have their means randomly located in the range $[-\pi,\pi]$. On the other hand, a lensed population will correspond to events with $\Delta\varphi_c = {0,\pi/4,\pi/2,3\pi/4}$. 

\paragraph{Analyzing different populations and detection thresholds.} To understand the effect of the assumptions of the population, we generate two other mock catalogs of unlensed events to be used with mass overlaps. For our second catalog, we use a different power-law mass distribution for the primary components mass: $p(m_1)\propto m_{1}^{-1.6}$ as opposed to $p(m_1)\propto m_{1}^{-2.35}$. 
For our third catalog, we test an extreme case in which the population has a prominent peak at a specific mass and described by a Gaussian:
\begin{equation}\label{Gaussian bump}
    p(m_1) = \frac{1}{\sigma \sqrt{2\pi}} \exp{-\frac{1}{2}\left( \frac{m_1 - \mu}{\sigma} \right)^2},
\end{equation}
where $\mu = 35\,M_{\odot}$, and $\sigma = 1$ (see the left panel of Fig.~\ref{fig:massOverlap_unlensed} for the primary component mass distribution of these catalogs). Moreover, we also analyze a more stringent network SNR condition ($\rho_{\rm obs}>24$) for both the mass overlap and sky map overlap to understand the effect of higher SNR thresholds on the false alarm probability.

\begin{figure*}[t!]
    \centering
    \includegraphics[width=\textwidth]{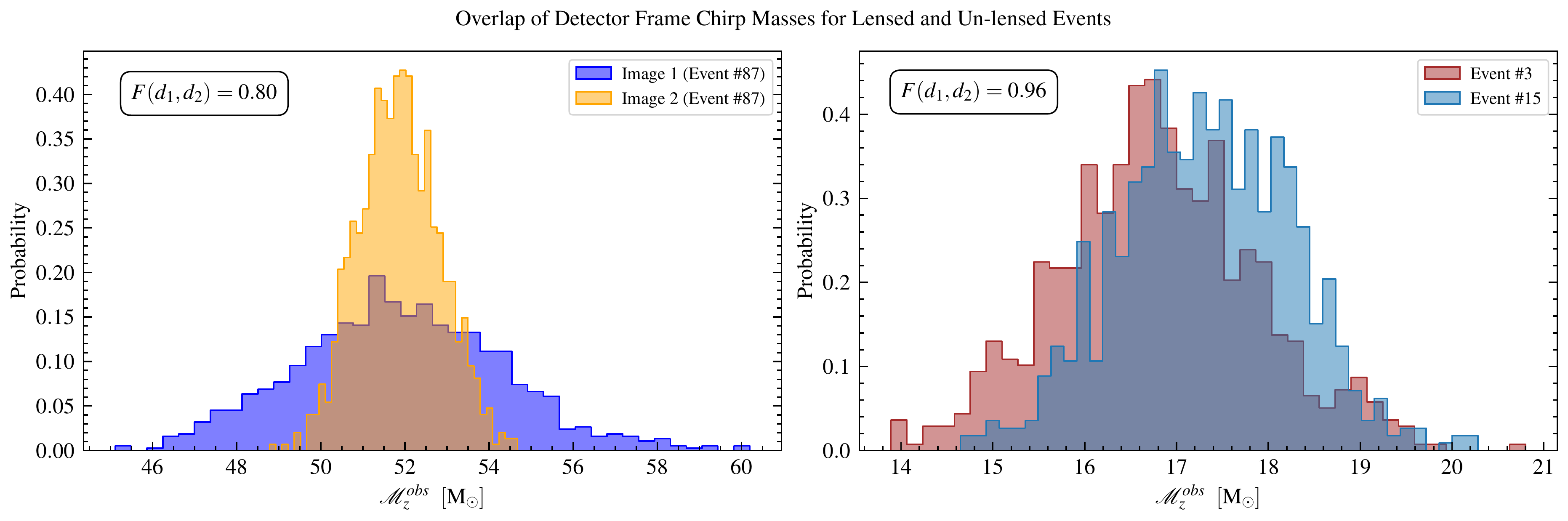}
    \caption{Probability density function of the posterior distribution of detector-frame chirp mass of a lensed pair (left) and an unlensed pair (right).
    The Bhattacharyya coefficients in this example, quantifying the amount of overlap between the pair for the lensed and the unlensed pairs, are $F=0.80$ and $F=0.96$, respectively. 
    The Bhattacharyya coefficient of the unlensed pairs can be high due to astronomical coincidence.
    For example, if only the detector-frame chirp mass were checked, this unlensed pair would appear more convincing that the true lensed pair, and thus lead to a false alarm.
    }
\label{fig:massOverlap_example}
\end{figure*}

%SUBSEC: PARAMETER OVERLAP
\subsection{Parameter overlap} \label{sec:parameter_Overlaps}

The product of GW parameter estimation of a given event is a multi-dimensional posterior distribution that describes the probability function of each of the parameters of the waveform model as well as their correlations (typically in 15 dimensions). 
Here, for simplicity, we will focus on reduced subsets of this parameter space. In particular, we concentrate on the parameters that are most relevant for the lensing hypothesis and better measured with current detectors: masses, sky localization, and phase.
We use a range of metrics to quantify the overlap between parameters:

\paragraph{Mass overlap.} For the masses, we quantify the parameter overlap between pairs of events using the Bhattacharyya coefficient\footnote{The Bhattacharyya coefficient $F$ is related to the Hellinger distance, $D_{H}$, a common statistical measure of the similarity between two probability distributions~\cite{Hellinger+1909+210+271}: $D_H(d_1,d_2) = \sqrt{1 - F(d_1, d_2)}$.}~\cite{Bhattacharyya}, $F$, 
also known as the fidelity~\cite{Vitale:2016jlv}. This is a statistical measure that quantifies the similarity of two distributions: 
\begin{equation}
    \begin{split}
    F(d_1,d_2)=&\int\sqrt{p(\Theta|d_1)p(\Theta|d_2)}d\Theta \\
    &=\left< \sqrt{\frac{p(\Theta|d_1)}{p(\Theta|d_2)}}\right>_{p(\Theta|d_2)}\,,
    \end{split}
\end{equation}
where $p(\Theta|d_1)$ and $p(\Theta|d_2)$ are the posterior distributions of the desired parameter, $\Theta$, for events one and two respectively.
Given that $p(\Theta|d_1)$ and $p(\Theta|d_2)$ are probability density functions, the Bhattacharyya coefficient is normalized within unity.

\paragraph{Sky map overlap.} For the sky maps, we quantify the parameter overlap between the pairs of events using
\begin{equation} \label{eq:skymap_overlap}
    \mathbb{O}(d_1,d_2) = \int p(\Omega|d_1)p(\Omega|d_2) \,d\Omega\,,
\end{equation}
where the integral is taken over all of the sky, and $p(\Omega|d_1)$ and $p(\Omega|d_2)$ are the posterior distributions of the parameter $\Omega$ (sky position) for events 1 and 2, respectively~\cite{Rico_overlap}.
As opposed to the Bhattacharyya coefficient, the sky map overlap $\mathbb{O}(d_1,d_2)$ is not normalized within unity. 
One normalization might be to divide the sky map overlap on the 2-D sky map posterior space by the area of the smaller sky map. 
However, if one of the sky maps is poorly constrained (e.g., $\Delta \Omega \approx 2000\ \rm deg^{2}$), we get an overlap value close to unity in many cases when the other sky map in the pair is better constrained and is enclosed by the poorly constrained sky map.
Instead, we enforce that all the sky maps have consistent pixel resolutions and use the overlap defined in Eq.~\ref{eq:skymap_overlap}.
This definition favors better-constrained sky maps (with smaller $\Delta \Omega$). 
For a more detailed discussion for different sky map overlap metrics, see~\cite{wong2021using}, and for a public version of the code used for this analysis, see~\cite{Rico_overlap}.

\paragraph{Phase overlap.} 
For the coalescence phase difference, we model the posterior distribution as a Gaussian. A normal distribution in periodic boundary conditions $[-\pi,\pi]$ can be described by an elliptic theta function
\begin{equation}
p(x,\mu,\sigma) = \frac{1}{2\pi}\vartheta_3 \left(\frac{x-\mu}{2},e^{-\sigma^2/2}\right).
\end{equation}
Therefore, to know the probability that the phase difference corresponds to the lensing prediction, we only need to compute $p(\mu=\Delta\varphi_c) = p(\Delta\varphi_c,\mu,\sigma) / p(\mu,\mu,\sigma)$, where the term in the denominator is the normalization factor.

%SUBSEC: FALSE ALARM PROBABILITY
\subsection{False alarm probability} \label{sec:fap}

A high Bhattacharyya coefficient and sky map overlap imply consistency with the lensing hypothesis. However, this is not equivalent to guaranteeing that an event is lensed.
Instead, we must consider the overlap in the context of the false alarm probability.

The total false alarm probability, $\rm FAP$, is the probability that, given a population of $N$ events, at least one pair within this population will mimic lensing due to astrophysical coincidence. Mathematically, the FAP is described by:
\begin{equation} \label{eq:totalFAP}
    \FAP  = 1-(1-\FAP_{\rm per\, pair})^{N_{\rm pairs}}\,,
\end{equation}
where $N_{\rm pairs}=N(N-1)/2$ is the number of unique pairs of events and $\FAP_{\rm per\, pair}$ (false alarm probability per pair) is the percentage of pairs with parameter overlaps similar to that of lensed pairs. In other words, $\FAP_{\rm per\, pair}$ is the percentage of pairs with parameter overlaps that are greater than or equal to a threshold overlap value defined by the behavior of parameter overlaps of the lensed pairs. 
Note that the \FAP\ and \FAPpair discussed here are different from the false alarm rate (FAR) used in GW search pipelines \cite{LIGOScientific:2016vbw}, which accounts for the trial factors under the noise background and quantifies the false alarm probability of the search pipeline.

To calculate the $\FAP_{\rm per\, pair}$, we first need to understand to what extent the parameters overlap for the lensed pairs. 
While the parameters of lens pairs exactly overlap in an ideal noiseless universe, the amount of overlap, in reality, varies due to detection errors and selection biases.
The SNRs of the lensed images of the same source will differ due to the magnification (de-magnification) and the difference in arrival time. Therefore, the posteriors of the detector-frame chirp mass and sky map will be slightly different for the lensed pairs. This means that, in the case of mass overlap, the Bhattacharyya coefficient for a lensed pair will not necessarily be 1. A similar situation is true for the overlap of other parameters.

For example, the Bhattacharyya coefficient for the unlensed pair in Fig.~\ref{fig:massOverlap_example} is 0.96, which is higher than the Bhattacharyya coefficient of 0.80 for the lensed pair. 
This shows explicitly how some unlensed pairs can mimic lensing due to coincidental parameter overlap.

To tackle the issue described above, we simulate a mock catalog of lensed events based on the description in Sec.~\ref{sec:mock_catalog}, and calculate the parameter overlaps based on Sec.~\ref{sec:parameter_Overlaps}.
We also calculate the cumulative distribution function of the parameter overlaps of the lensed pairs. This allows us to understand the threshold overlap amount, which is the amount of overlap surpassed by most of the lensed pairs. For example, we look for the amount of overlap displayed by the upper 99\%, 95\%, and 50\% of the lensed pairs ($\rm 1^{st}$ and $\rm 5^{th}$ percentiles, and the median of the CDF, respectively).
We then look for the percentage of unlensed pairs with an overlap value greater than the threshold, finding the $\FAP_{\rm per\, pair}$ for that parameter for the catalog of unlensed events.

It is an arbitrary choice to pick the overlap threshold.
Accepting the $\rm 1^{st}$ or $\rm 5^{th}$ percentile as the threshold accounts for most of the lensed pairs but also leads to a greater \FAP\ since the threshold value is lower. Setting the threshold higher would lead to a lower \FAP\ at the cost of missing more lensed pairs.

The lensing hypothesis suggests that all the parameters we focus on (mass, sky map, and shift in the coalescence phase) should overlap. 
Hence, significant overlap in one of the parameters is not sufficient to identify lensing. After finding $\FAP_{\rm per\, pair}$ for the unlensed population, we calculate the probability that all three parameters overlap due to coincidence. 
Then, given a population of $N$ events, we can finally calculate the total false alarm probability for the catalog, $\FAP$. 
Although there could be weak correlations between these three parameters, we conservatively study them independently and compute the total $\FAP_{\rm per\, pair}$ as the product of each of them.

%--------
%SECTION: RESULTS
%--------
\section{Results} \label{sec:results}

In the following, we present our results for the overlaps of mass, sky map, and coalescence phase. We then compute the total false alarm probability. 

%SUBSEC: MASS OVERLAPS
\subsection{Mass overlap} \label{sec:mass_overlap}

\begin{figure}[htp]
    \centering
    \includegraphics[width=\columnwidth]{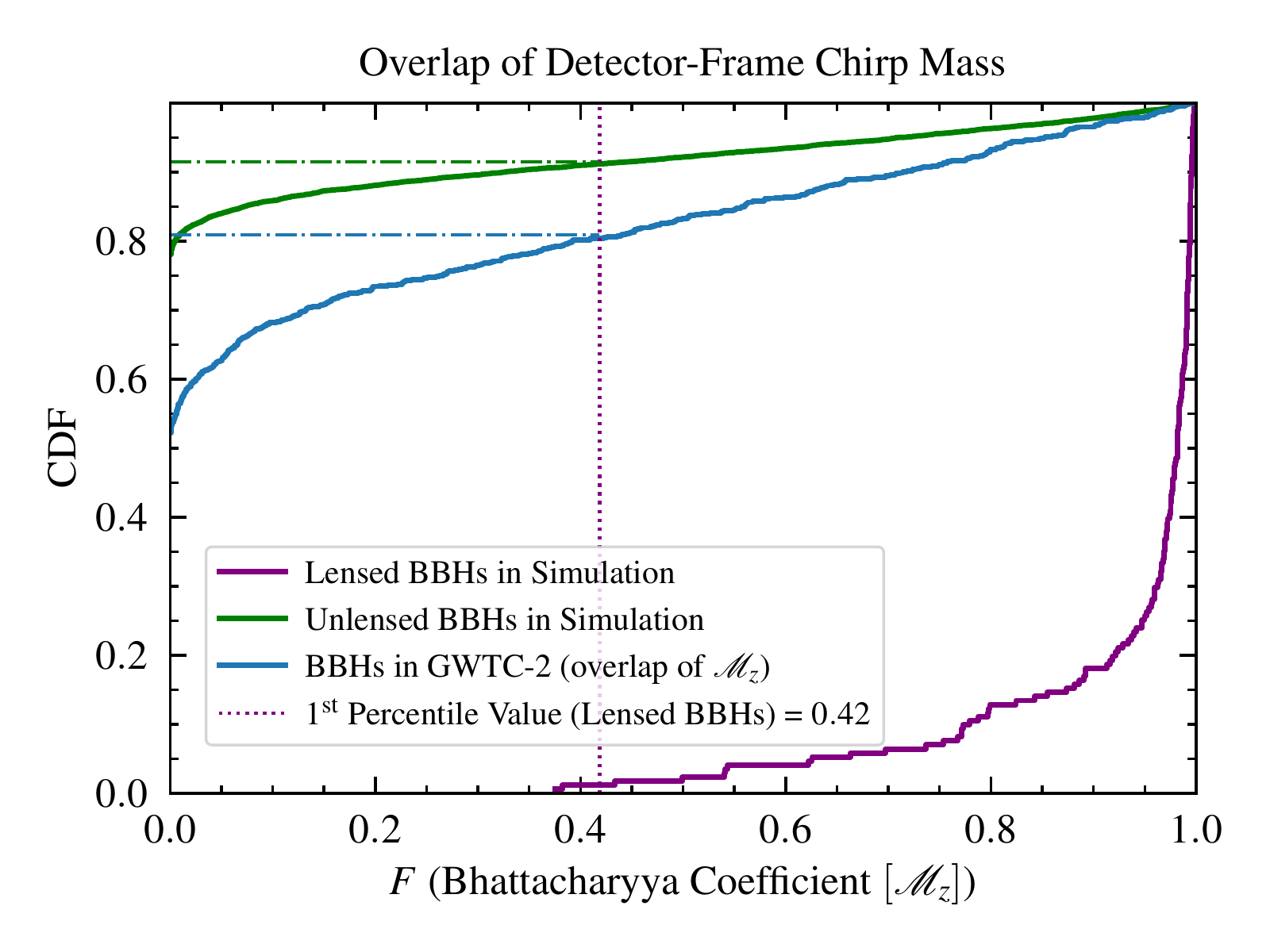}
    \caption{
    Cumulative distribution function of the Bhattacharyya coefficient $F$ based on the detector-frame chirp mass for simulated lensed (purple) and unlensed (green) binary black holes (BBHs), as well as BBHs found in GWTC-2~\citep{Abbott_2019, Abbott_2021} (blue). 
    We can use the cumulative distribution to assess the lensing false alarm probability.
    Although the lensed BBH pairs produce significantly higher Bhattacharyya coefficients ($F>0.42$ at 99\% confidence), the unlensed pairs still mimic lensing ($F>0.42$ in 9\% and 20\% of the simulated and catalog pairs, respectively). 
    These unlensed pairs with high Bhattacharyya coefficients lead to false alarms.
    We have assumed a power-law mass distribution for the primary component mass of the BBHs, and use the SDSS galaxy catalog with a singular isothermal ellipsoid model for the lenses.
    }
    \label{fig:combined_mass}
\end{figure}

We begin by computing the overlap in the masses. Given that GW detectors are sensitive to the detector-frame chirp mass ($\mathcal{M}_z$) at the leading post-Newtonian order, to simplify our analysis, we focus on this quantity instead of the source frame component masses; they are related by Eq.~\ref{eq:ChirpMass}. 
This choice allows us to avoid the bias in the inferred source frame component masses of lensed images that arises from the magnification of the signal.
The results of the overlap of $\mathcal{M}_z$ for our simulated population of lensed and unlensed events are presented in Fig.~\ref{fig:combined_mass}, where we show the cumulative distribution function (CDF) of the Bhattacharyya coefficient $F$. 

In addition, we compute the distribution of $F$ for the BBH mergers reported in the catalog GWTC-2~\citep{Abbott_2019, Abbott_2021} encompassing the observing runs O1, O2, and O3a. 
We perform this computation both for $\mathcal{M}_z$ and $M^{\rm total}_{z}$ to understand whether either one performs better in terms of false alarm probability. The result for the overlap of $\mathcal{M}_z$ is also presented in Fig.~\ref{fig:combined_mass}. Since we find that $\mathcal{M}_z$ and $M^{\rm total}_{z}$ perform similarly in terms of false alarm probability per pair, the result for the overlap of $M^{\rm total}_{z}$ is omitted from Fig. ~\ref{fig:combined_mass}.

\begin{figure*}[t]
    \centering
    \includegraphics[width=\textwidth]{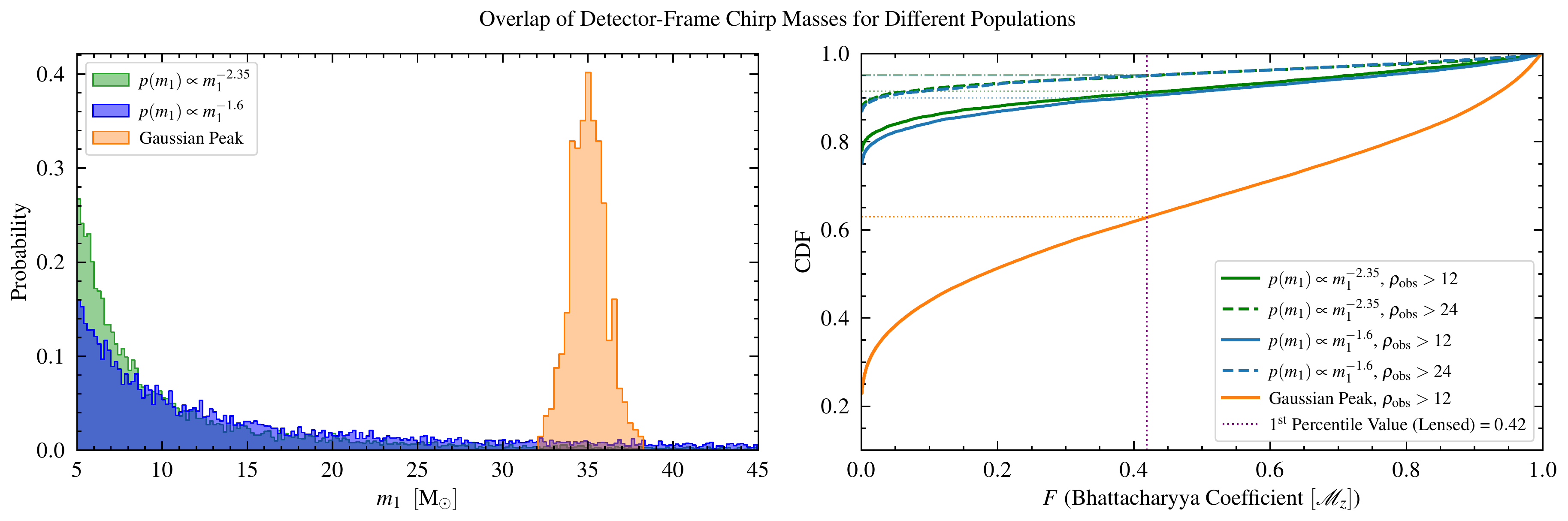}
    \caption{Simulated BBH populations with different mass distributions (left) and the cumulative distribution function of the Bhattacharyya coefficient based on the overlap of detector-frame chirp mass for these populations (right).
    We investigate three different distributions for the primary component mass of the BBHs: $p(m_1)\propto m_{1}^{-2.35}$ (green), $p(m_1)\propto m_{1}^{-1.6}$ (blue), and an extreme distribution for which the primary component mass is distributed as a Gaussian (orange).
    The secondary component mass is uniformly distributed in the mass ratio in each case.
    We also investigate the effect of using a more stringent signal-to-noise ratio (SNR) threshold on the false alarm probability per pair: $\rho_{\rm obs}>12$ (solid lines on the right) and $\rho_{\rm obs}>24$ (dashed lines on the right).
    In each case, unlensed pairs still mimic lensing ($F>0.42$ in 9\% and 10\% of the simulated pairs with $p(m_1)\propto m_{1}^{-2.35}$ and $p(m_1)\propto m_{1}^{-1.6}$, respectively).
    Doubling the SNR threshold ($\rho_{\rm obs}>24$) lowers the false alarm probability per pair ($F>0.42$ in 5\% of the simulated pairs with $p(m_1)\propto m_{1}^{-2.35}$ and $p(m_1)\propto m_{1}^{-1.6}$).
    For the distribution with the Gaussian peak, $F>0.42$ in 37\% of the simulated pairs, showing that if there was a formation channel that created binaries with similar masses, the false alarm probability would increase (a few factors) due to the increased overlap of chirp mass.
    }
\label{fig:massOverlap_unlensed}
\end{figure*}

The purple line in Fig.~\ref{fig:combined_mass} shows the CDF of $F$ for the lensed BBH pairs in our simulation.
The $\rm 1^{st}$ percentile, the $\rm 5^{th}$ percentile, and the median values of $F$ (corresponding to the upper $99\%$, $95\%$ and $50\%$) for the lensed population are $F_{0.01}^{\rm lensed} = 0.52$, $F_{0.05}^{\rm lensed} = 0.87$, and $F_{0.5}^{\rm lensed} = 0.98$, respectively.
We show the threshold based on the upper 99\% of the lensed pairs ($\rm 1^{st}$ percentile value) with the purple dotted vertical line.
The percentage of unlensed pairs above this threshold can be found by finding the CDF value of the point where the vertical (threshold) dotted line intersects the CDF for the unlensed population and subtracting this value from unity.
For example, the dotted line intersects the unlensed population in the simulation approximately at 0.91 (in the vertical axis). This means $\sim 9\%$ of the unlensed pairs produce Bhattacharyya coefficients greater than this threshold, while $99\%$ of the true lensed pairs are above this threshold.

Even though the lensed pairs produce significantly higher Bhattacharyya coefficients ($F>0.42$ at 99\% confidence), the unlensed pairs still mimic lensing ($F>0.42$ in 9\% and 20\% of the simulated (green) and catalog (blue) pairs, respectively). This means that if we wanted to detect 99\% of the lensed pairs, approximately 9\% of the unlensed pairs would also mimic lensing if only the mass overlap is considered. Therefore, \FAPpair $\ = 9\%$ (based on mass overlap).

As discussed in Sec.~\ref{sec:fap}, the threshold value could be set considering different fractions of the lensed pairs.
We present the results based on different thresholds in Table~\ref{tab:mass_results}. 
As we increase the overlap threshold, fewer unlensed events mimic lensing at the cost of losing some lensed events from the catalog.

BBHs in GWTC-2 produce a distribution very similar to the unlensed BBHs in our simulation. The \FAPpair based on mass overlap is slightly lower for the simulated unlensed BBHs than the BBHs in GWTC-2 because for the simulation we assume design sensitivity. Larger errors increase the probability of overlap. 
The comparison of relative uncertainty of $\mathcal{M}_z$ between the mock catalog and the BBHs in GWTC-2 can be found in Appendix~\ref{app:mock_vs_real}. We find that the relative uncertainty behaves differently for $\mathcal{M}_z$ and $M^{\rm total}_{z}$ at different mass ranges (low mass vs. high mass). In particular, low mass events have better detector-frame chirp mass measurements.
We also provide comparisons between the SNR distribution of the mock catalog of unlensed and lensed events and the BBHs in GWTC-2 in Appendix~\ref{app:comparison_of_SNR}, finding an overall consistency.

If we consider a more stringent SNR threshold, e.g.~$\rho_{\rm obs} > 24$, then a smaller portion of the unlensed BBHs ($\sim5\%$) have Bhattacharyya coefficients as large as those of the lensed BBHs. This lowers the number of false alarms and indicates that the events with higher SNRs are better candidates for the conclusive detection of lensing. Moreover, one could also focus on the low mass part of the population. This is because the relative uncertainty of the detector-frame chirp mass is better constrained for low mass events as opposed to high mass events (for a detailed discussion, see Appendix~\ref{app:low_vs_high}). Therefore, accidental overlap due to errors in the parameter estimation is less likely to happen, reducing the false alarm probability.

\paragraph{Effects of different mass distributions and more stringent SNR thresholds.} We compute the overlap in detector-frame chirp masses for different assumptions for the mass distribution of the unlensed population and analyze the effect of applying a more stringent SNR threshold, $\rho_{\rm obs} > 24$, on the $\text{FAP}_{\text{per pair}}$ as described in Sec.~\ref{sec:mock_catalog}. 

In Fig.~\ref{fig:massOverlap_unlensed}, we present the probability distribution function for the primary component mass of the BBHs based on different populations on the left side. The green and the blue represent two power-law distributions with different indices ($\alpha = -2.35$ and $\alpha = -1.6$, respectively). The orange represents the case in which the primary component mass distribution is a Gaussian,  as described in Sec.~\ref{sec:mock_catalog}. This is an extreme, limiting case motivated by the observation of a peak in the mass distribution at $\sim35\ \rm M_\odot$ \cite{LIGOScientific:2021psn}, which is consistent with the prediction of pulsational pair instability supernova theory \cite{Talbot:2018cva}.
On the right side of Fig.~\ref{fig:massOverlap_unlensed}, we present the CDF of the Bhattacharyya coefficient similarly to Fig.~\ref{fig:combined_mass}. Here, the dashed lines represent the cases in which we apply the $\rho_{\rm obs} > 24$ condition.

In each case, the unlensed pairs still mimic lensing ($F>0.42$ in 9\% and 10\% of the simulated pairs with $p(m_1)\propto m_{1}^{-2.35}$ and $p(m_1)\propto m_{1}^{-1.6}$, respectively). Doubling the SNR threshold ($\rho_{\rm obs}>24$) lowers the false alarm probability per pair ($F>0.42$ in 5\% of the simulated pairs with $p(m_1)\propto m_{1}^{-2.35}$ and $p(m_1)\propto m_{1}^{-1.6}$). For the distribution with the Gaussian peak, $F>0.42$ in 37\% of the simulated pairs, implying that if the underlying mass distribution has peaks at given mass values, the FAP would increase considerably due to the increased mass overlap.
Although we do not consider spin in this analysis, uncertainties in the spin would similarly contribute to the number of false alarms.
Thus the underlying distribution of masses can significantly alter the false alarm probabilities.
We summarize the results based on different thresholds for these populations in Table~\ref{tab:mass_results}.

%%%                           %%%
%%% MASS RESULT SUMMARY TABLE %%%
%%%                           %%%
\begin{table}[ht]
\caption{Summary of the \FAPpair based on mass overlaps of unlensed populations for different assumed mass distributions and different overlap thresholds.}\vspace{2mm}
\setlength{\extrarowheight}{2pt}
\centering
\begin{adjustbox}{width=1\columnwidth}
\begin{tabular}{lccc}
\hline
\multicolumn{1}{c}{\multirow{2}{*}{Unlensed Population}} & \multicolumn{3}{c}{\FAPpair} \\ \cline{2-4} 
\multicolumn{1}{c}{} &
  \begin{tabular}[c]{@{}c@{}}99\% \\ ($1^{\rm st}$ percentile)\end{tabular} &
  \begin{tabular}[c]{@{}c@{}}95\% \\ ($5^{\rm th}$ percentile)\end{tabular} &
  \begin{tabular}[c]{@{}c@{}}50\% \\ (median)\end{tabular} \\ \hline
$p(m_1)\propto m_{1}^{-2.35}$                            & 0.09   & 0.06  & 0.005  \\
$p(m_1)\propto m_{1}^{-2.35}\ (\rho_{\rm obs}>24)$       & 0.05   & 0.03  & 0.004  \\
$p(m_1)\propto m_{1}^{-1.6}$                             & 0.10   & 0.07  & 0.006  \\
$p(m_1)\propto m_{1}^{-1.6}\ (\rho_{\rm obs}>24)$        & 0.05    & 0.03  & 0.004  \\
Gaussian Peak                                            & 0.37   & 0.27  & 0.03   \\
BBHs in GWTC-2                                           & 0.20   & 0.12  & 0.005  \\ \hline
\end{tabular}
\end{adjustbox}
\label{tab:mass_results}
\end{table}

%SUBSEC: SKY MAPS OVERLAPS
\subsection{Sky map overlap} \label{sec:skyMap_overlap}

\begin{figure}[b]
    \centering
    \includegraphics[width=\columnwidth]{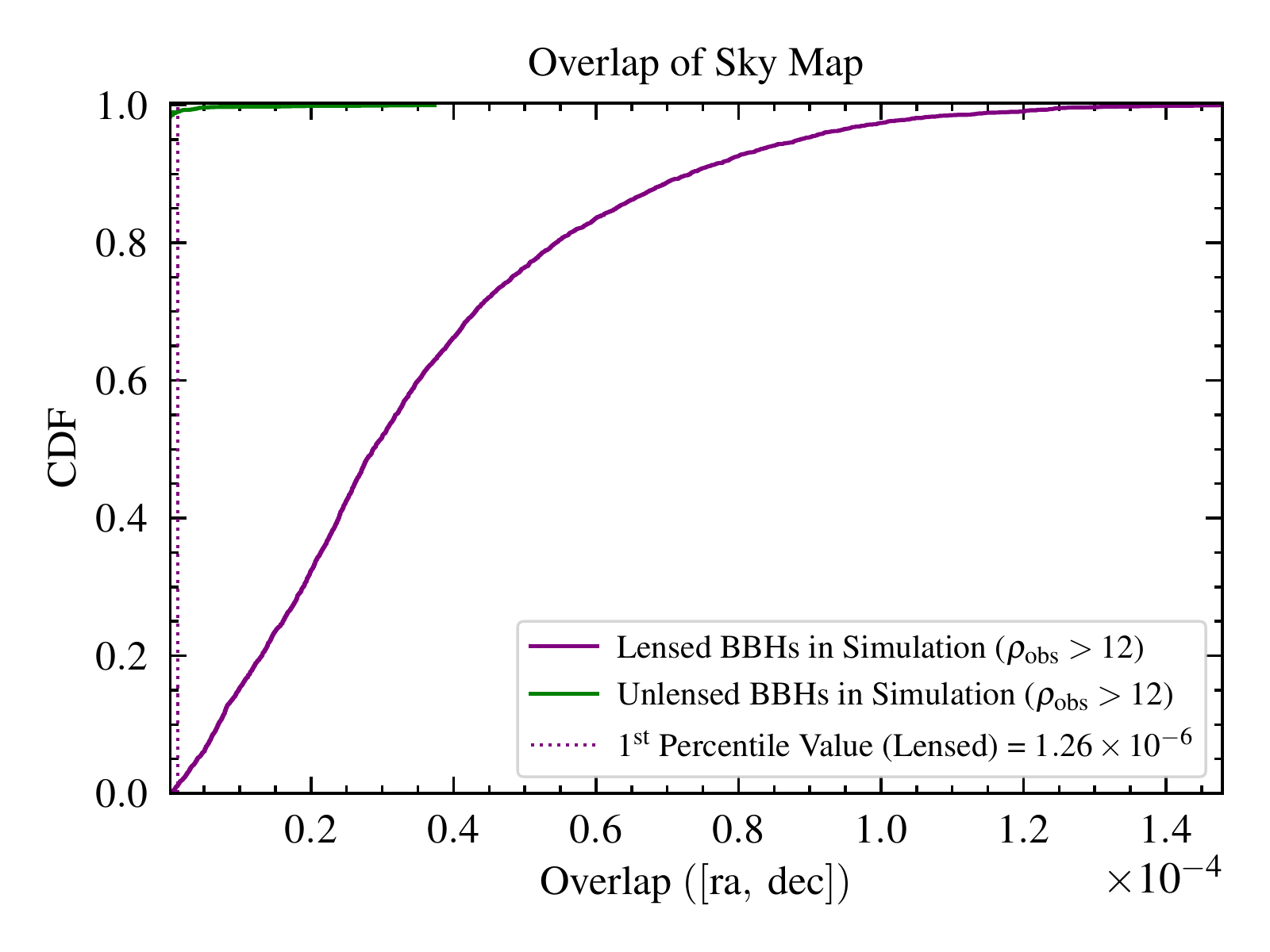}
    \caption{
        Cumulative distribution function of the overlap of sky maps of simulated lensed (purple) and unlensed (green) binaries. 
        Although the lensed BBH pairs produce significantly higher sky map overlaps ($\mathbb{O} > 1.26 \times 10^{-6}$ at 99\% confidence), the unlensed pairs can still mimic lensing ($\mathbb{O} > 1.26 \times 10^{-6}$ in 1\% of the simulated pairs).
        These unlensed pairs with high sky map overlaps lead to false alarms.
        We generate and localize these lensed and unlensed events using \texttt{BAYESTAR}~\citep{PhysRevD.93.024013}.
        }
    \label{fig:skyMap_CDF_SNR16}
\end{figure}

We continue our analysis by computing the overlap in the sky maps, $\mathbb{O}(d_1,d_2)$. We present the results for our simulated population of lensed and unlensed events in Fig.~\ref{fig:skyMap_CDF_SNR16}.

\begin{figure*}[t!]
    \centering
    \subfloat[\centering Lensed pair]{ {\includegraphics[width=7cm]{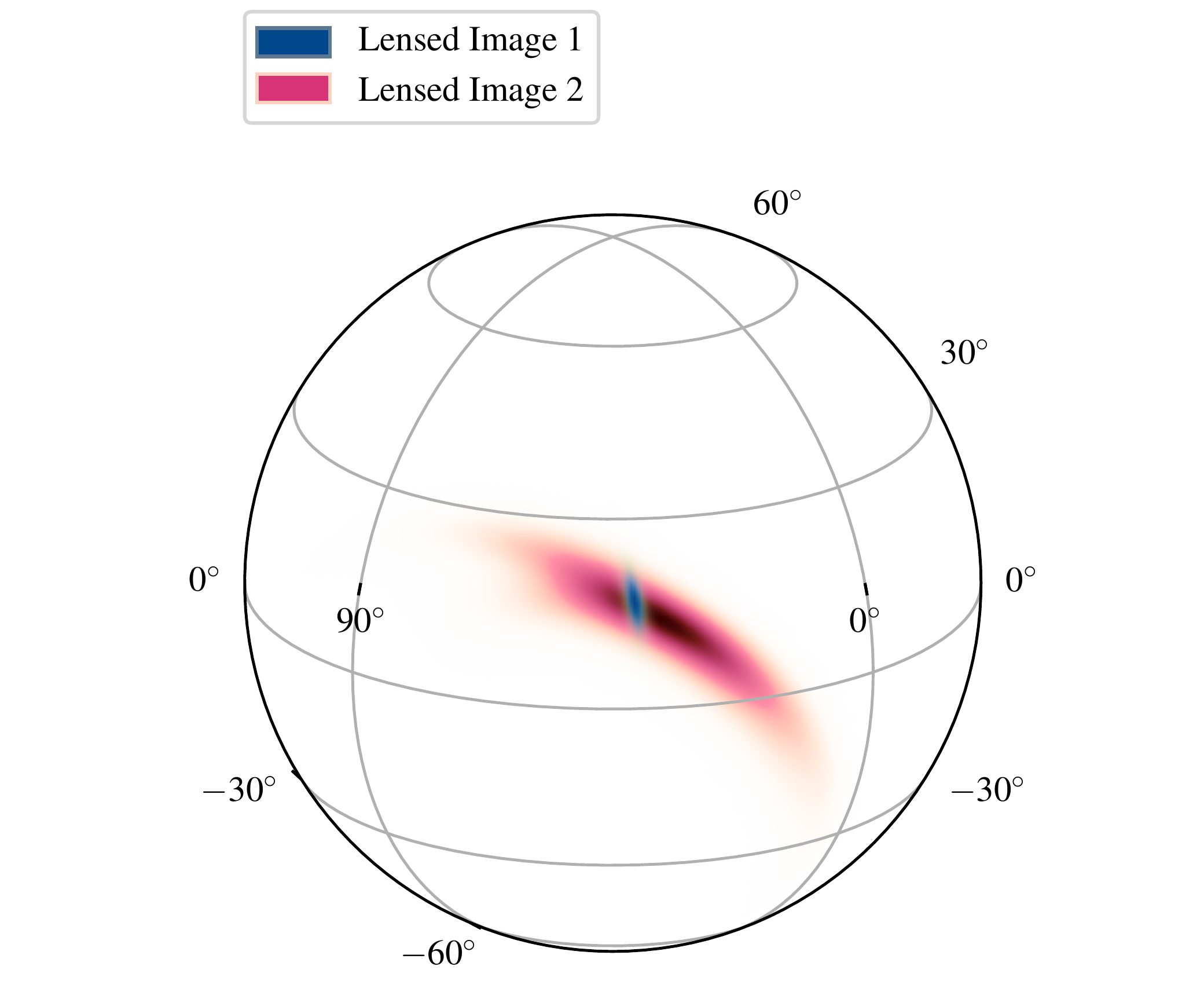} }}%
    \qquad
    \centering
    \subfloat[\centering Unlensed pair]{ {\includegraphics[width=7cm]{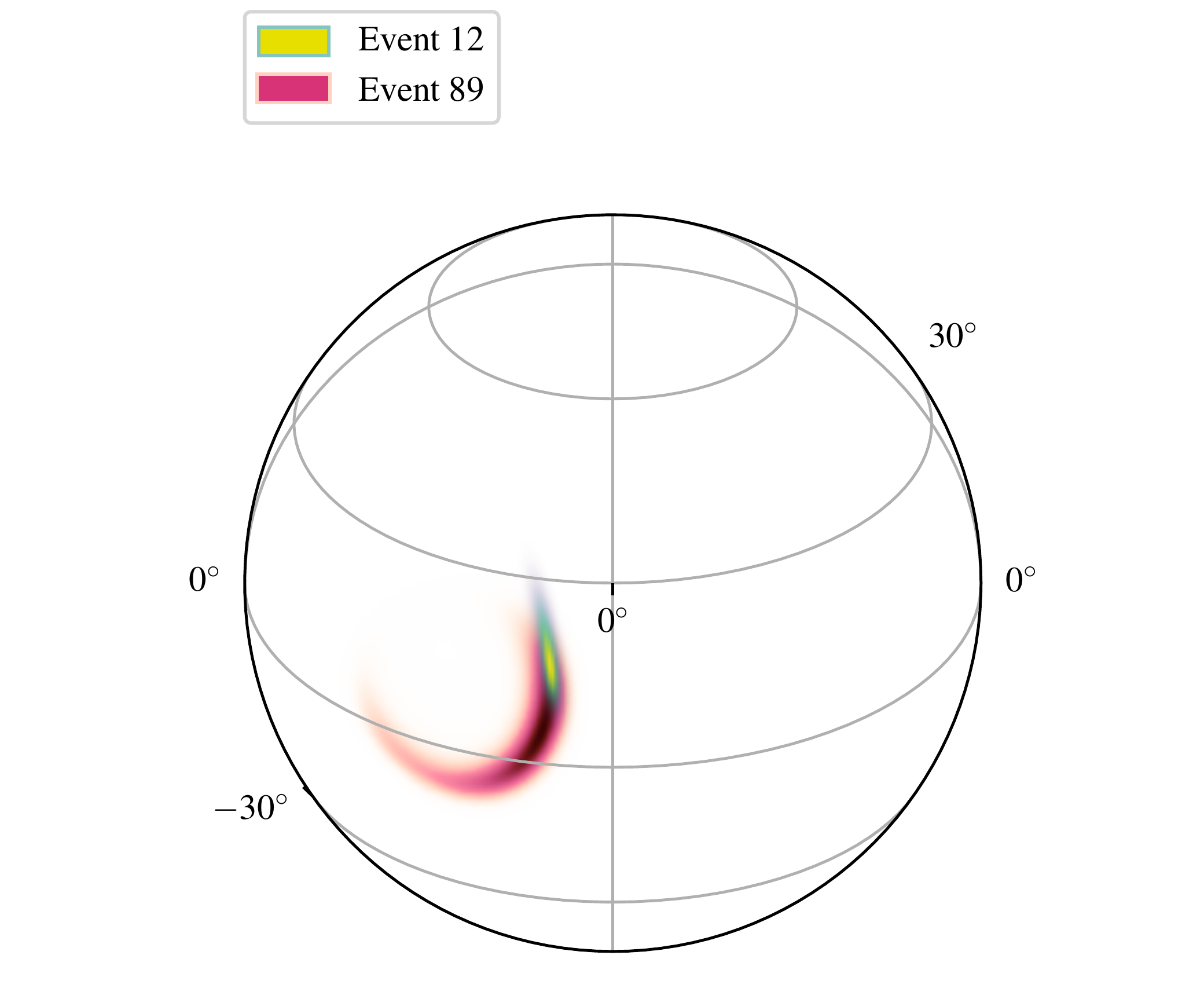} }}%
    \caption{Examples of possible sky maps of a lensed pair (left) and two unlensed events (right). The sky map overlaps of the lensed and unlensed pairs are $\mathbb{O}_{\rm lensed} = 8.28 \times 10^{-5}$ and $\mathbb{O}_{\rm unlensed} = 2.72 \times 10^{-5}$, respectively.
    Even though the overlap of the lensed pair is higher, the unlensed pair pair also exhibit a high amount of overlap (above the $95\%$ threshold), mimicking lensing.}%
    \label{fig:skymapOverlap_example}%
\end{figure*}

The purple line in Fig.~\ref{fig:skyMap_CDF_SNR16} shows the CDF of $\mathbb{O}(d_1,d_2)$ for the lensed BBHs in our simulation. The $\rm 1^{st}$ percentile, the $\rm 5^{th}$ percentile, and the median values of $\mathbb{O}(d_1,d_2)$ for the lensed population are $\mathbb{O}_{0.01}^{\rm lensed} = 1.26 \times 10^{-6}$, $\mathbb{O}_{0.05}^{\rm lensed} = 4.11 \times 10^{-6}$, and $\mathbb{O}_{0.5}^{\rm lensed} = 2.88 \times 10^{-5}$, respectively. Similarly to Figs.~\ref{fig:combined_mass} and~\ref{fig:massOverlap_unlensed}, we show the threshold based on the upper 99\% of the lensed pairs with the purple dotted vertical line.

Even though the lensed BBH pairs produce significantly higher sky map overlaps ($\mathbb{O} > 1.26 \times 10^{-6}$ at 99\% confidence), the unlensed pairs can still mimic lensing ($\mathbb{O} > 1.26 \times 10^{-6}$ in 1\% of the simulated pairs).
These unlensed pairs with high sky map overlaps lead to false alarms.
Increasing the network SNR threshold $\rho_{\rm obs}$ from 12 to 24 improves the localization, resulting in a significant decrease in the false alarm probability per pair ($\mathbb{O} > 1.26 \times 10^{-6}$ in 0.2\% of the simulated pairs with $\rho_{\rm obs}>24$). 
We present the results based on different thresholds in Table~\ref{tab:skyMap_results}, and an illustrative example of sky maps of lensed and unlensed pairs can be seen in Fig.~\ref{fig:skymapOverlap_example}.
An alternative to the high SNR cut would be to focus on three-detector detections since their localization is highly improved.

%%%                           %%%
%%% SKY MAP RESULT SUMMARY TABLE %%%
%%%                           %%%
\begin{table}[ht]
\caption{Summary of \FAPpair based on sky map overlaps of different network SNR thresholds.}\vspace{2mm}
\setlength{\extrarowheight}{2pt}
\centering
\begin{adjustbox}{width=1\columnwidth}
\begin{tabular}{lccc}
\hline
\multicolumn{1}{c}{\multirow{2}{*}{Network SNR Threshold}} & \multicolumn{3}{c}{\FAPpair} \\ \cline{2-4} 
\multicolumn{1}{c}{} &
  \begin{tabular}[c]{@{}c@{}}99\% \\ ($1^{\rm st}$ percentile)\end{tabular} &
  \begin{tabular}[c]{@{}c@{}}95\% \\ ($5^{\rm th}$ percentile)\end{tabular} &
  \begin{tabular}[c]{@{}c@{}}50\% \\ (median)\end{tabular} \\ \hline
$\rho_{\rm obs}>12$                                      & 0.01  & 0.005  & $6 \times 10^{-4}$  \\
$\rho_{\rm obs}>24$                                      & 0.002 & $9 \times 10^{-4}$  & $2 \times 10^{-4}$ \\ \hline
\end{tabular}
\end{adjustbox}
\label{tab:skyMap_results}
\end{table}

%SUBSEC: COALESCENCE PHASE OVERLAPS
\subsection{Coalescence phase overlap} \label{sec:phase_overlap}

\begin{figure*}[t!]
    \centering
    \includegraphics[width=\textwidth]{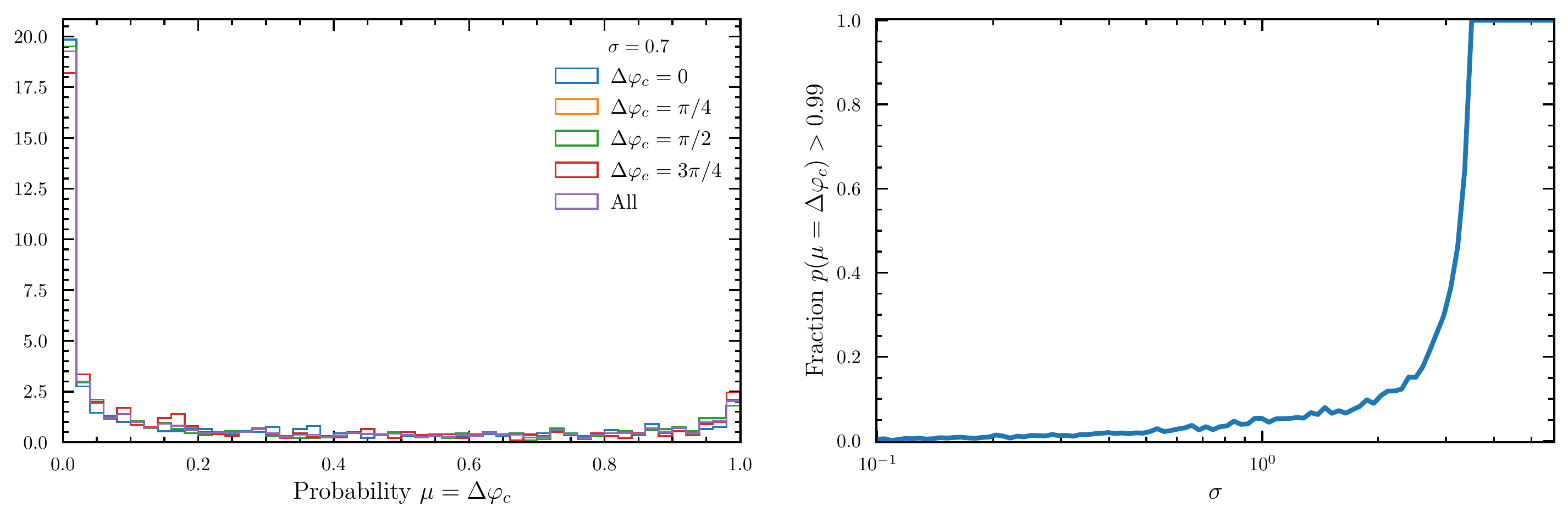}
    \caption{On the left, probability that the coalescence phase difference posterior is centered around the strong lensing prediction $\mu=0,\pi/4,\pi/2,3\pi/4$ for a fixed value of the dispersion of the normal distribution $\sigma=0.7$. On the right, fraction of events with a probability of $\mu=\Delta\varphi_c$ larger than $99\%$ as a function of the dispersion. For example, for $\sigma=0.7$ as in \cite{Dai:2020tpj}, the fraction of events is $10\%$.}
    \label{fig:phase_overlap}
\end{figure*}

Finally, we compute the probability that the coalescence phase difference coincides with the strong lensing prediction. 
Our results are given in Fig. \ref{fig:phase_overlap}. First, we simulate a set of 1000 coalescence phase difference posteriors with fixed dispersion and random mean values. We choose $\sigma=0.7$ following the joint PE analysis of \cite{Dai:2020tpj}. For each of these simulations we compute the probability of $\mu=\Delta\varphi_c$ for the lensing phase shifts $\Delta\varphi_c=0,\pi/4,\pi/2,3\pi/4$. This corresponds to the left panel of Fig. \ref{fig:phase_overlap}. In this case, $10\%$ of the events will have a probability of coincidence with lensing larger than $99\%$. 
This is a rough, conservative value and more detailed analyses are necessary.

Since the phase difference dispersion value is the crucial parameter in this calculation, we repeat the above exercise for different values from 0.1 to $2\pi$. As plotted in the right panel of Fig.~\ref{fig:phase_overlap}, there is a rapid increase of mimicking the lensing prediction as a function of $\sigma$, with a better-measured phase reducing the phase FAP. In Appendix~\ref{app:measured_phase}, we discuss a more adequate choice of the phase that is designed to match the actual phase measured by a GW detector. A more detailed analysis will be granted in future work.

As an alternative to the phase overlap, one could use the fact that type II lensed images may exhibit waveform distortions to identify them as strongly lensed \cite{Ezquiaga:2020gdt}. However, these distortions are subdominant unless higher-order modes are well measured, or the signal is very loud. Therefore, these types of searches may be better suited for next-generation detectors \cite{Wang:2021kzt}, although if a strongly lensed event is observed in conjuction with higher-order modes, a joint analysis of the two signals may reveal the small distortions~\citep{Lo:2021nae,Janquart:2021nus}.

\subsection{Total False Alarm Probability} \label{sec:total_FAP}

We compute the false alarm probability per pair, \FAPpair, based on the description in Sec.~\ref{sec:fap}. Using the threshold set by the overlaps produced by the upper 99\% of the lensed pairs, the \FAPpair (for the simultaneous overlap of mass, sky map, and shift in coalescence phase) is given by
\begin{equation} \label{eq:FAPpair}
\begin{split}
\rm FAP_{per\ pair} & \approx 0.09 \times 0.01 \times 0.10 \\
 & \approx 10^{-4}.
\end{split}
\end{equation}
This implies that given $10^{4}$ unlensed pairs, approximately $1$ of them will mimic lensing and cause a false alarm.

When we consider the \FAPpair based on the same threshold but with the more stringent SNR condition ($\rho_{\rm obs}>24$), we find 
\begin{equation} \label{eq:FAPpair_SNR16}
\begin{split}
\rm FAP_{per\ pair}^{\rho_{\rm obs}>24} & \approx 0.05 \times 0.002 \times 0.1 \\
 & \approx 10^{-5},
\end{split}
\end{equation}
which means that if we only considered pairs of observations for which $\rho_{\rm obs}>24$, then only $1$ in $10^{5}$ of these (unlensed) pairs would be lensing false alarms. However, as a drawback, we would be missing a large fraction of the true lensed events if we set $\rho_{\rm obs}>24$. We report the results of \FAPpair for the other populations and assumptions in Table~\ref{tab:combinedResults}.

We calculate the total false alarm probability \FAP\ based on Eq.~\ref{eq:totalFAP} using the \FAPpair value in Eq.~\ref{eq:FAPpair}.
We plot the result as a function of the total number of detections in Fig.~\ref{fig:FAP_vs_events}. 
Our results indicate that if we detected a total of 100 events, the probability of having at least one false alarm in this event set is already approximately 1 ($\rm FAP\approx 1$). Assuming the lensing rate is $10^{-3}$, then when we have a total of $10^{3}$ events, we would expect to see multiple false alarms even though we would expect to see only $1$ lensed event. Therefore, we may observe multiple false alarms in the future if we base our analysis on waveform and sky map overlap.

%%%                           %%%
%%% FAP RESULT SUMMARY TABLE  %%%
%%%                           %%%

\begin{table}[ht]
\caption{Summary of \FAPpair for mass, sky map, coalescence phase, and the simultaneous overlap of all the parameters. We assume that \FAPpair based on coalescence phase shift is constant for different thresholds.
We use the network SNR threshold $\rho_{\rm thr} = 12$. For  high $\rho_{\rm thr}$, we fix the network SNR threshold to 24.}\vspace{2mm}
\setlength{\extrarowheight}{2pt}
\centering
\begin{adjustbox}{width=1\columnwidth}
\begin{tabular}{lccc}
\hline
\multicolumn{1}{c}{\multirow{2}{*}{Parameter}} & \multicolumn{3}{c}{\FAPpair}                                      \\ \cline{2-4} 
\multicolumn{1}{c}{} &
  \begin{tabular}[c]{@{}c@{}}99\% \\ ($1^{\rm st}$ percentile)\end{tabular} &
  \begin{tabular}[c]{@{}c@{}}95\% \\ ($5^{\rm th}$ percentile)\end{tabular} &
  \begin{tabular}[c]{@{}c@{}}50\% \\ (median)\end{tabular} \\ \hline
Mass                                           & 0.09               & 0.06               & 0.005              \\
Mass (high $\rho_{\rm thr}$)                     & 0.05               & 0.03               & 0.004              \\
Sky Map                   & 0.01               & 0.005              & $6 \times 10^{-4}$              \\
Sky Map (high $\rho_{\rm thr}$)                  & 0.002           & $9 \times 10^{-4}$               & $2 \times 10^{-4}$               \\
Coalescence Phase                              & 0.1                & 0.1                & 0.1                \\ \hline
Combined                                       & $1 \times 10^{-4}$ & $3 \times 10^{-5}$ & $3 \times 10^{-7}$ \\ 
Combined (high $\rho_{\rm thr}$)                 & $1 \times 10^{-5}$ & $3 \times 10^{-6}$ & $8 \times 10^{-8}$ \\ \hline
\end{tabular}
\end{adjustbox}
\label{tab:combinedResults}
\end{table}

\subsection{Effect of triple and quadruple images and higher signal-to-noise ratio thresholds}

Some lensed BBHs may be triply or quadruply-imaged depending on the source-lens configuration. We roughly estimate \FAPtriple\ and \FAPquadruple\ (the FAP per triplet and the FAP per quadruplet) using the results of $\rm FAP_{per\ pair}$. While the probability of a coincidental overlap of three or four events is significantly lower (compared to only two events), the number of lensed triplets and quadruplets is also much lower.

\begin{figure}[h]
    \centering
    \includegraphics[width=\columnwidth]{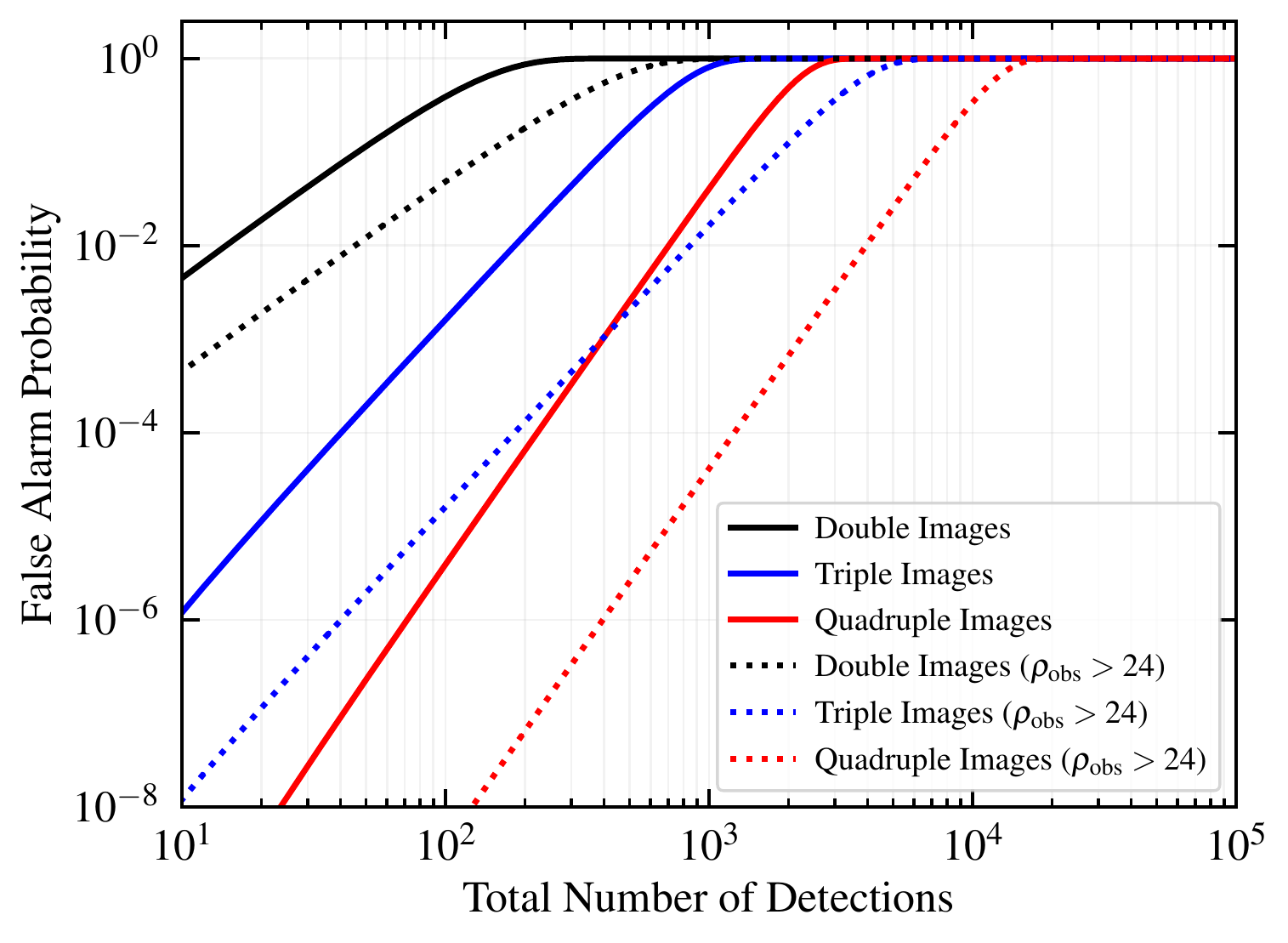}
    \caption{
        False alarm probability as a function of the total number of BBH mergers in the catalog. The black, blue, and red lines represent the FAP for double, triple, and quadruple images, respectively. 
        The dotted lines represent the more stringent case of $\rho_{\rm obs}>24$ for their respective image numbers. 
        For the double and triple images, the false alarm probability is approximately 1 once the total event number is $\mathcal{O}(1000)$.
        When we consider more stringent cases, such as quadruple images with $\rho_{\rm obs}>24$, the false alarm probability is still approximately 1 once the event number is high enough to expect a quadruply lensed event with $\rho_{\rm obs}>24$.
        This means that when the total number of events is high enough to expect lensing to be detected, the probability of having false alarms is already comparably high in each case.
        }
    \label{fig:FAP_vs_events}
\end{figure}

We calculate the FAP using Eq.~\ref{eq:totalFAP}, but this time we use $N_{\rm triple}$ and $N_{\rm quadruple}$, the number of triplets and quadruples, instead of $N_{\rm pair}$.
We plot the results as a function of the total number of detections (including the $\rho_{\rm obs}>24$ condition)  in Fig.~\ref{fig:FAP_vs_events}. 

Even though the \FAP\ is lower for triply-imaged events (and even lower for the quadruply-imaged events), we find $\rm FAP \approx 1$ for any case in consideration when we reach $\mathcal{O}(1000)$ total events (corresponding to $\sim10^6$ unique pairs). Even for the rarest cases, such as quadruple images with $\rho_{\rm obs}>24$, we expect to see at least one false alarm once the total number of detections is high enough to expect a quadruply lensed event with $\rho_{\rm obs}>24$. Therefore, when the total number of events is high enough to expect lensing to be present in the sample, the probability of having a false alarm due to chance alignment is already greater.

The reason that the false alarms inevitably dominate over the true lensed populations is that the former grows much faster than the latter. In particular, the total FAP for lensing (for double images) is
\begin{equation}
    \FAP \propto N^2\ \FAP_{\rm per \, pair}\,,
\end{equation}
where $N$ is the number of observed events.
In other words, it becomes increasingly likely that at least one pair of events have overlapping detector-frame parameters and sky maps as we observe more GW events. The expected number of lensed events, on the other hand, is proportional to the total number of detected events, $\propto N$.

When we investigate triply and quadruply-imaged events, which have lower false alarm probabilities for individual detections compared to doubly-imaged events, this issue becomes more dominant since
\begin{equation}
    \FAP \propto N^3\ \FAP_{\rm per \, triplet}\,
\end{equation}
when we investigate triply-imaged events, and FAP scales $\propto N^4$ for quadruply-imaged events.
In other words, although it is less likely for any three or four random unlensed events to have overlapping detector-frame parameters and sky location, given a set of $N$ events, there are $\mathcal{O}(N^3)$ triplets and $\mathcal{O}(N^4)$ quadruples, meaning that the overall \FAP\ will increase more rapidly with an increasing number of events.
In conclusion, as the number of events increases, the number of lensing false alarms will increase more rapidly than the number of actual lensed events in our sample. Without improved lensed-event selection, the true lenses will be lost in a sea of false ones.

\subsection{Conclusive detection of lensing}

Building upon our previous results, we investigate the possibility of the conclusive detection of strong lensing of GWs, i.e., constructing a search such that the number of lensed events is larger than the expected number of false alarms.
To be agnostic about the lens model, we calculate the expected number of lensed events and false alarms as a function of the total number of detections for three different lensing rates: $10^{-2}$, $10^{-3}$, and $10^{-4}$~\citep{Xu:2021bfn}.
We repeat this calculation for two SNR thresholds, $\rho_{\rm obs}>12$ and $\rho_{\rm obs}>24$, and consider both the $1^{\rm st}$ percentile (upper $99\%$) and median (upper $50\%$)  thresholds. For each case, we use the \FAPpair values presented in Table~\ref{tab:combinedResults}.

Using the higher SNR threshold, $\rho_{\rm thr}^{\rm new}$, significantly lowers the $\rm FAP_{per\ pair}$ at the cost of decreasing the sample size (by $\sim (\rho_{\rm thr}/\rho_{\rm thr}^{\rm new})^3$), and, therefore, reducing the number of lensed events in the sample by the same factor.
A similar reduction occurs when we choose a stricter lensing threshold.

\begin{figure*}[t!]
    \centering
    \includegraphics[width=\textwidth]{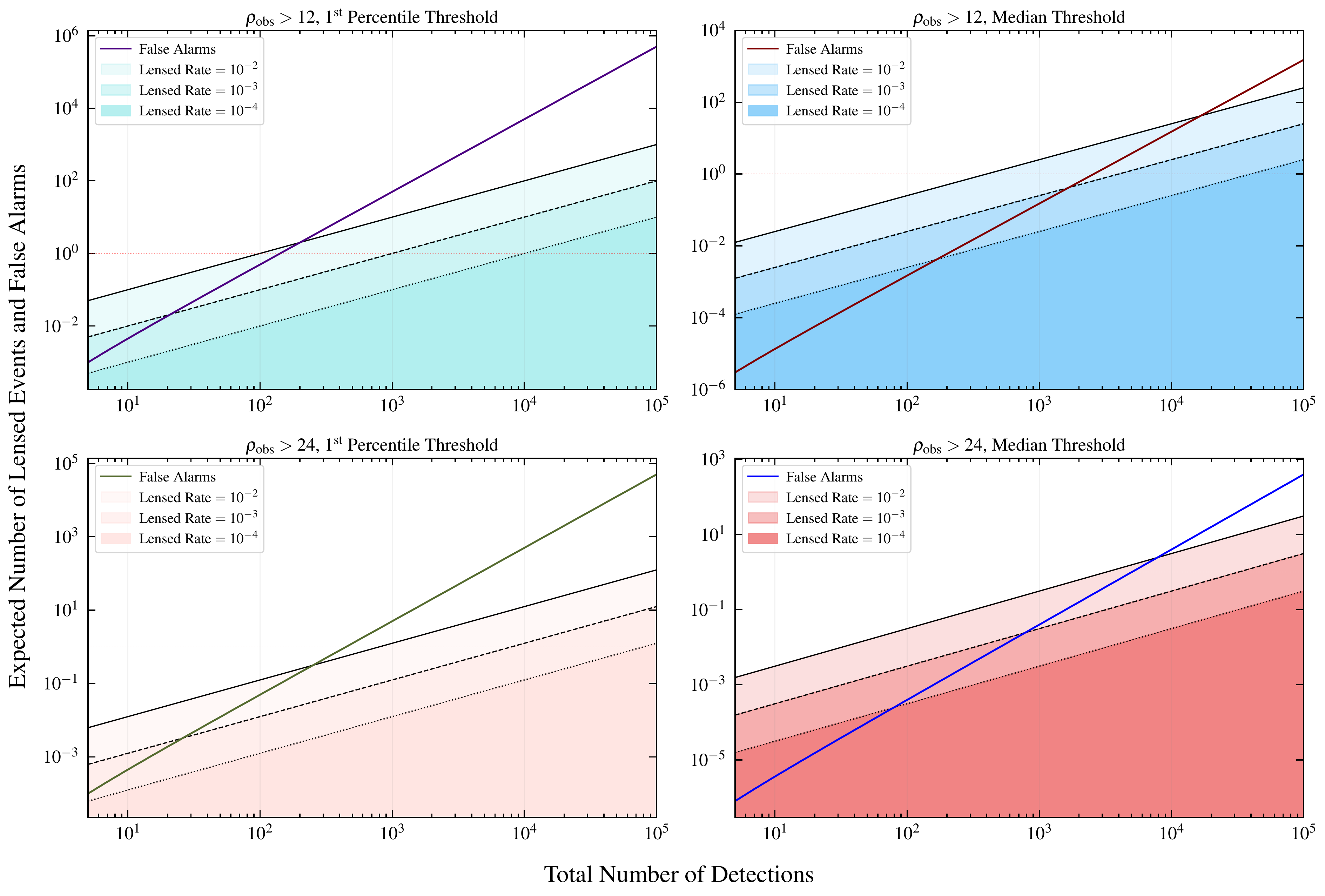}
    \caption{Expected number of lensed events and false alarms as a function of the total number of detections for different thresholds in terms of the amount of overlap (1st percentile vs. median) as well as signal-to-noise ratio (SNR) cuts.
    \textit{Upper left:} overlap threshold is based on the $1^{\rm st}$ percentile value (upper 99\%)  of the lensed pairs; SNR cut is $\rho_{\rm obs} > 12$. 
    \textit{Upper right:} overlap threshold is based on the median value (upper 50\%)  of the lensed pairs; SNR cut is $\rho_{\rm obs} > 12$.
    \textit{Lower left:} overlap threshold is based on the $1^{\rm st}$ percentile value; SNR cut is $\rho_{\rm obs} > 24$.
    \textit{Lower right:} overlap threshold is based on the median value; SNR cut is $\rho_{\rm obs} > 24$. 
    Considering the number of detections needed to expect one lensed event for each lensing rate, the only lensing rate that wins over the false alarms is $10^{-2}$. 
    For other lensing rates, when the number of events is high enough to expect one lensed event, the expected number of false alarms is already higher.
    }
\label{fig:Exp_FA_vs_Lensed}
\end{figure*}

We present our results in Fig.~\ref{fig:Exp_FA_vs_Lensed}.
We find that the number of true lensed events dominates over the false alarms only for a lensing rate of
$10^{-2}$, a rate that is highly disfavored by the current data~\cite{Buscicchio:2020cij,LIGOScientific:2021izm,Xu:2021bfn}. 
For other lensing rates, when the number of events is high enough to expect at least one lensed event in the sample, there are already a greater number of expected false alarms.

This indicates that with our simple overlap criteria for current detectors at design sensitivity, the false alarms will win over realistic lensing rates ($\lesssim10^{-3}$) even when selecting the highest SNR pairs. This result highlights the necessity to design alternative identification criteria beyond simple waveform and sky location overlap for the conclusive detection of strong lensing.

One could improve this issue by incorporating the time-delay prior of lensing (effectively time windowing), which can be obtained from electromagnetic lensing measurements~\citep{Haris:2018vmn,Wierda:2021upe,More:2021kpb}. 
The vast majority of galaxy lenses create time delays of around minutes to months (with lower time-delay being favored), while the galaxy cluster lensing time delays could be of the order of years.
Therefore, we could preferentially focus on events separated by no less than a few minutes and no more than a few years---alternatively, we could directly incorporate a full time-delay prior to our search. This approach would reduce the number of relevant pairs to consider for possible lensing.
We might achieve a similar reduction of the FAP by introducing prior information on the magnification distribution. 
However, in such cases, it will become important to ensure that the results will not be dominated by uncertainties in the time-delay or magnification distributions.

Another possible improvement could be provided by the next-generation (3G) GW detectors, such as Cosmic Explorer \cite{Evans:2021gyd} and Einstein Telescope \cite{Maggiore:2019uih}. 
The improved parameter estimation, especially the precise sky localization, could decrease the degree of coincidental overlap of parameters, and significantly lower the FAP.
Since these detectors are also expected to measure other parameters such as spin better, one could consider these other parameters in addition to the overlap of mass, sky map, and phase.
This could further decrease the number of unlensed pairs mimicking lensing.

One could also use the waveform distortions exhibited by type II lensed images to identify lensed events. 
Since these distortions are not dominant unless higher-order modes, precession, or eccentricity are well measured~\cite{Ezquiaga:2020gdt}, the next-generation detectors could utilize searches involving these waveform distortions to establish conclusive detection of lensing.

Moreover, since these next-generation detectors are expected to detect a higher number of events each year, the increased number of events could enable us to place higher thresholds on the lensing candidates. We could select a higher SNR threshold or only consider candidates with high amounts of parameter/localization overlap and do so without lowering the occurrence of lensing in the sample to a negligible rate.
However, the high number of observed events with the next-generation detectors leads to significantly higher false alarm probabilities since these increase as $\sim N^2$. We leave a detailed study of lensing false alarms for the next-generation detectors for future work.

Even though the parameter overlap might not provide a conclusive detection of lensing, we can use a subset of the candidate pairs that show high amounts of overlap as triggers for electromagnetic searches. These electromagnetic searches could look for lensing galaxies that match the expected lens mass. Furthermore, if a BBH is strongly lensed, its host galaxy could also be strongly lensed by the same lensing galaxy. Therefore, these electromagnetic searches could also utilize the magnification ratios between the lensed images inferred using the GW detectors and look for lensed background galaxies that have multiple images with the same magnification ratios. One could perform cosmography studies if the actual lensing galaxy and the host galaxy are found.~\cite{Hannuksela:2020xor}

\begin{figure}[h]
    \centering
    \includegraphics[width=\columnwidth]{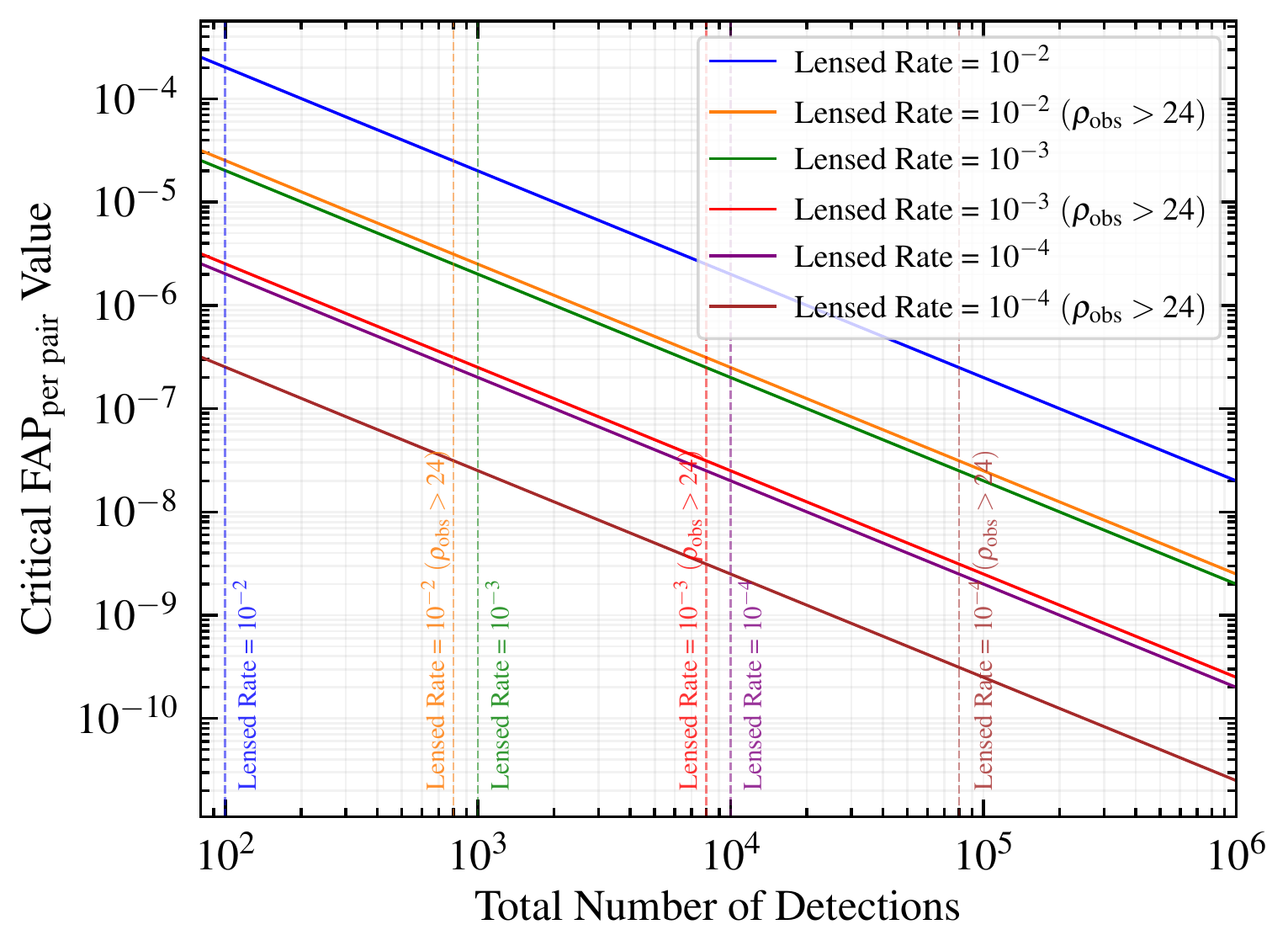}
    \caption{Critical value of false alarm probability per pair, $\rm FAP_{per\ pair}$, as a function of the total number of detections in the GW transient catalog, $N$. This value represents how low the \FAPpair should be for the expected number of lensed events to be the same as the expected number of false alarms for a given $N$. We present the result for three lensing rates: $10^{-2}$, $10^{-3}$, and $10^{-4}$. For each of these lensing rates, we also consider the case in which we select a higher signal-to-noise threshold, $\rho_{\rm obs}> 24$. We indicate the total number of detections for which we expect one lensed event with the vertical dashed lines for each lensing rate. For the number of lensed events to be higher than the number of false alarms in the sample, the \FAPpair value should be lower than the critical value shown in the plot.
        }
    \label{fig:criticalFAP}
\end{figure}

To estimate how low the \FAPpair should be so that the expected number of lensed events is higher than the expected number of false alarms, we analytically calculate the critical \FAPpair as a function of the number of detections, $N$, for different lensing rates and SNR thresholds. 
This follows directly from the definition of the FAP in Eq.~(\ref{eq:totalFAP}). 
We present the results in Fig.~\ref{fig:criticalFAP}. We find that, if the lensing rate is $10^{-3}$, then the \FAPpair should be $\lesssim 2\times10^{-6}$ so that the expected number of false alarms is not higher than the expected number of lensed events (which is 1 if the total number of detections is 1000). If the \FAPpair value is higher than this, we will have multiple false alarms while we may only have one lensed event, increasing the difficulty of confidently distinguishing the true lensing event from the false ones.

%--------
%SECTION: CONCLUSIONS
%--------
\section{Conclusions} \label{sec:conclusions}

Strong lensing of GWs is a unique probe of the matter distribution in the universe and a magnifying glass to study the properties of the population of compact objects \cite{Xu:2021bfn}. However, definitively distinguishing pairs of lensed sources from random associations presents complex data analysis problems. As the number of events, $N$, in the GW catalogs increases, the number of pairs of events increases as $\sim N^2$. This means that the probability of having unlensed events which mimic the parameter overlap in mass and sky localization expected from strong lensing also increases as $\sim N^2$, thereby leading to false alarms. Meanwhile, the number of lensed events will increase linearly with $N$, implying that for sufficiently high $N$, the false alarms will always dominate over the actual lensing events.

We have constructed mock catalogs of lensed and unlensed events and investigated the degree to which unlensed events mimic lensed ones because of the overlap of parameters due to a combination of random coincidence and errors in the parameter estimation.
We find that the probability of a false alarm based on coincidental overlap of the chirp mass, sky location, and coalescence phase are approximately $9\%$, $1\%$, and $10\%$ per pair, respectively.
Combining the three, we arrive at a false alarm probability per pair of $\sim10^{-4}$.
We validate our simulation against the GWTC-2 data, finding that the catalog data is consistent with our expectations.

We find that if we consider a more stringent SNR threshold, e.g., $\rho_{\rm obs} > 24$, a smaller portion of the unlensed BBHs mimics lensing due to improved parameter estimation. 
In this case, we find that the probability of a false alarm based on coincidental overlaps of chirp mass and sky location are approximately 5\% and 0.2\% per pair, respectively. 
As a result, we find that the combined false alarm probability per pair for this SNR threshold is $\sim10^{-5}$.

We summarize our results for the mass overlap in Figs.~\ref{fig:combined_mass} and~\ref{fig:massOverlap_unlensed} and Table~\ref{tab:mass_results}, our results for the sky map overlap in Fig.~\ref{fig:skyMap_CDF_SNR16} and Table~\ref{tab:skyMap_results}, our results for the phase overlap in Fig.~\ref{fig:phase_overlap}, and our combined results for all the parameters and all the different overlap and SNR thresholds under consideration in Table~\ref{tab:combinedResults}. 

We find that the relative uncertainty of detector-frame chirp mass is lower for low mass events than high mass events. As a result, the \FAPpair based on the overlap of mass is lower if the mass distribution of the BBHs has more low mass events as opposed to high mass events. Moreover, we analyze the effects of introducing a Gaussian peak to the distribution of the primary component mass of the BBHs and find that in this case \FAPpair based on mass increases to $\sim 37\%$ (as compared to $\sim 9\%$ without the Gaussian peak). 
This result shows that if there was a formation channel that created binaries with a peak in mass, the false alarm probability might significantly increase due to the increased mass overlap. 
Although we do not consider the spin in this analysis, uncertainties in the spin would similarly contribute to the number of false alarms. 
Indeed, the number of false alarms depends sensitively on the population of binaries in the Universe.

We show that sky map overlaps perform better than mass and phase in terms of distinguishing lensed events. 
We find that using stricter overlap thresholds (based on the lensed pairs that show a higher amount of overlap) and having higher SNR cuts decreases the $\rm FAP_{per\ pair}$. In turn, the overall \FAP\ also decreases. 
However, looking for a higher amount of overlap reduces the number of unlensed pairs that mimic lensing at the cost of missing some of the true lensed pairs ($\propto \rho_{\rm thr}^{-3}$).

Based on our results, we compute the false alarm probability, i.e., the probability of having at least one false alarm, as a function of $N$. We present our results in Fig.~\ref{fig:FAP_vs_events}. We find that if we detect $\mathcal{O}(100)$ events, the \FAP\ approaches unity. Assuming that the lensing rate is $\lesssim 10^{-3}$, when we have a total of $10^{3}$ events, we would expect to see multiple false alarms even though the expected number of lensed events would be 1. This suggests that we may face multiple false alarms in the near future and that their number will dominate over the real lensed events. 

Strongly lensed events could lead to more than two detectable images. Therefore,  
we estimate the \FAP\ for triply and quadruply-imaged events. We find that the \FAPtriple\ and \FAPquadruple\ are considerably lower than $\rm FAP_{per\ pair}$. 
However, the number of possible triplets and quadruplets in the catalog rapidly increases due to the larger number of possible combinations. As a result, the \FAP\ for triply and quadruply-imaged events reaches unity before the total number of detections is high enough to expect one true lensed event of this sort, as we show in Fig.~\ref{fig:FAP_vs_events}.

We summarize our main results in Fig.~\ref{fig:Exp_FA_vs_Lensed}, where we estimate when there will be a conclusive detection of strong lensing, i.e., when the number of true lensing detections outnumbers the false alarms. 
We find that for realistic lensing rates ($\lesssim10^{-3}$), current detectors at design sensitivity (observing thousands of events per year) will be dominated by false alarms. 
In addition, we analytically calculate how low the \FAPpair should be in order to have more lensed events than false alarms for a given number of detections in the GW catalog (Fig.~\ref{fig:criticalFAP}).

The results of this work demonstrate the necessity of designing more robust identification criteria, beyond simple binary parameter and sky location overlap, to identify lensed events. These are particularly important for next-generation detectors such as Cosmic Explorer and Einstein Telescope, for which hundreds of strongly lensed events are expected per year. 
We discuss several possible improvements, including sharper SNR cuts, improved parameter estimation, time delay priors, and looking for distortions due to the lensing phase. 
Our methods and results for the computation of FAP from an astrophysical population of BBHs could be extrapolated to the search of repeated GW signals from triple systems \cite{Veske:2020idq} or echoes in modified gravity \cite{Ezquiaga:2021ler}.

\acknowledgements

We are grateful to Neha Anil Kumar, Roberto Cotesta, Emanuele Berti, and Marc Kamionkowski for helpful comments on this manuscript. 
We are thankful to Rico K. L. Lo, Ken Ng, Luca Reali, and Javier Roulet for valuable discussions. We also thank Rico K. L. Lo for his guidance in using his sky map overlap calculation code. 
We are grateful to Justin Janquart for input on the expected error on the Morse phase. 
M\c{C} is supported by NSF Grants No. PHY-1818899, PHY-1912550 and AST-2006538, NASA ATP Grants No. 17-ATP17-0225 and 19-ATP19-0051, NSF-XSEDE Grant No. PHY-090003, NSF Grant PHY-20043, and Johns Hopkins University through the Rowland Research Fellowship. JME is supported by NASA through the NASA Hubble Fellowship grant HST-HF2-51435.001-A awarded by the Space Telescope Science Institute, which is operated by the Association of Universities for Research in Astronomy, Inc., for NASA, under contract NAS5-26555. M\c{C}, JME, and DEH are supported by NSF grants PHY-2006645, PHY-2011997, and PHY-2110507, as well as by the Kavli Institute for Cosmological Physics through an endowment from the Kavli Foundation and its founder Fred Kavli. OAH was partially supported by grants from the Research Grants Council of the Hong Kong (Project No. CUHK 14306218), The Croucher Foundation of Hong Kong and Research Committee of the Chinese University of Hong Kong.
DEH also gratefully acknowledges the Marion and Stuart Rice Award. 
This material is based upon work supported by NSF LIGO Laboratory which is a major facility fully funded by the National Science Foundation. 
This research project was conducted using computational resources at the Maryland Advanced Research Computing Center (MARCC).

\appendix

%---------
%APPENDIX: COMPARISON WITH REAL EVENTS
%---------
\section{Comparing the simulations with real events} \label{app:mock_vs_real}

In this section, we compare our simulations with the BBH mergers in GWTC-2~\citep{Abbott_2019, Abbott_2021} as a sanity check of our results. In Sec.~\ref{app:low_vs_high}, we analyze the relative uncertainty of chirp mass and total mass posteriors for low versus high mass events to understand whether using one parameter over the other could be more beneficial in terms of the false alarm probability depending on how high (or low) the chirp mass/total mass is. In Sec.~\ref{app:comparison_of_uncertainties}, we compare the relative uncertainty in the chirp mass posteriors of the unlensed BBHs in simulation with those of the BBHs in GWTC-2. In Sec.~\ref{app:comparison_of_SNR}, we compare the SNR of both the unlensed and lensed BBHs in simulation with those of the BBHs in GWTC-2.

\subsection{Relative uncertainty of chirp mass and total mass for low vs. high mass events} \label{app:low_vs_high}

\begin{figure}[htp]
    \centering
    \includegraphics[width=\columnwidth]{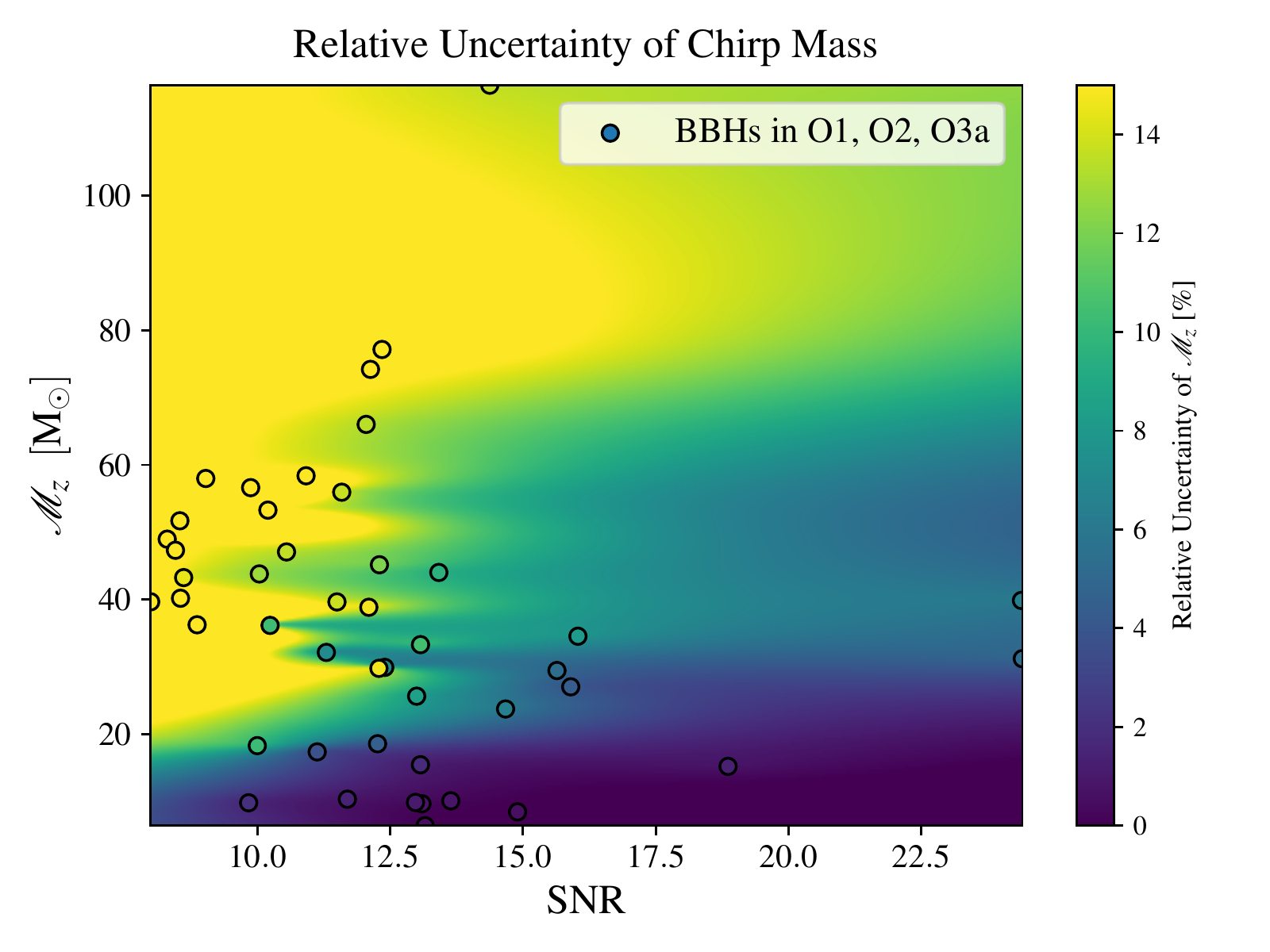}
    \caption{
    Contour map of relative uncertainty of detector-frame chirp mass as a function of detector-frame chirp mass and SNR. \textit{Horizontal-axis:} the SNR of the event. \textit{Vertical-axis:} the detector-frame chirp mass of the event. \textit{Right-bar:} relative uncertainty of the event (based on $95\%$ confidence interval). Data points are BBHs in GWTC-2~\citep{Abbott_2019, Abbott_2021}. The color of the data points represents the actual relative uncertainty for the specific event. The rest of the contour plot is completed by interpolation. Any point with relative uncertainty higher than $15\%$ is also represented by the yellow color. Lower-mass events generally have a better-constrained chirp mass than high-mass events.
    }
    \label{fig:contour_chirp}
\end{figure}

\begin{figure}[htp]
    \centering
    \includegraphics[width=\columnwidth]{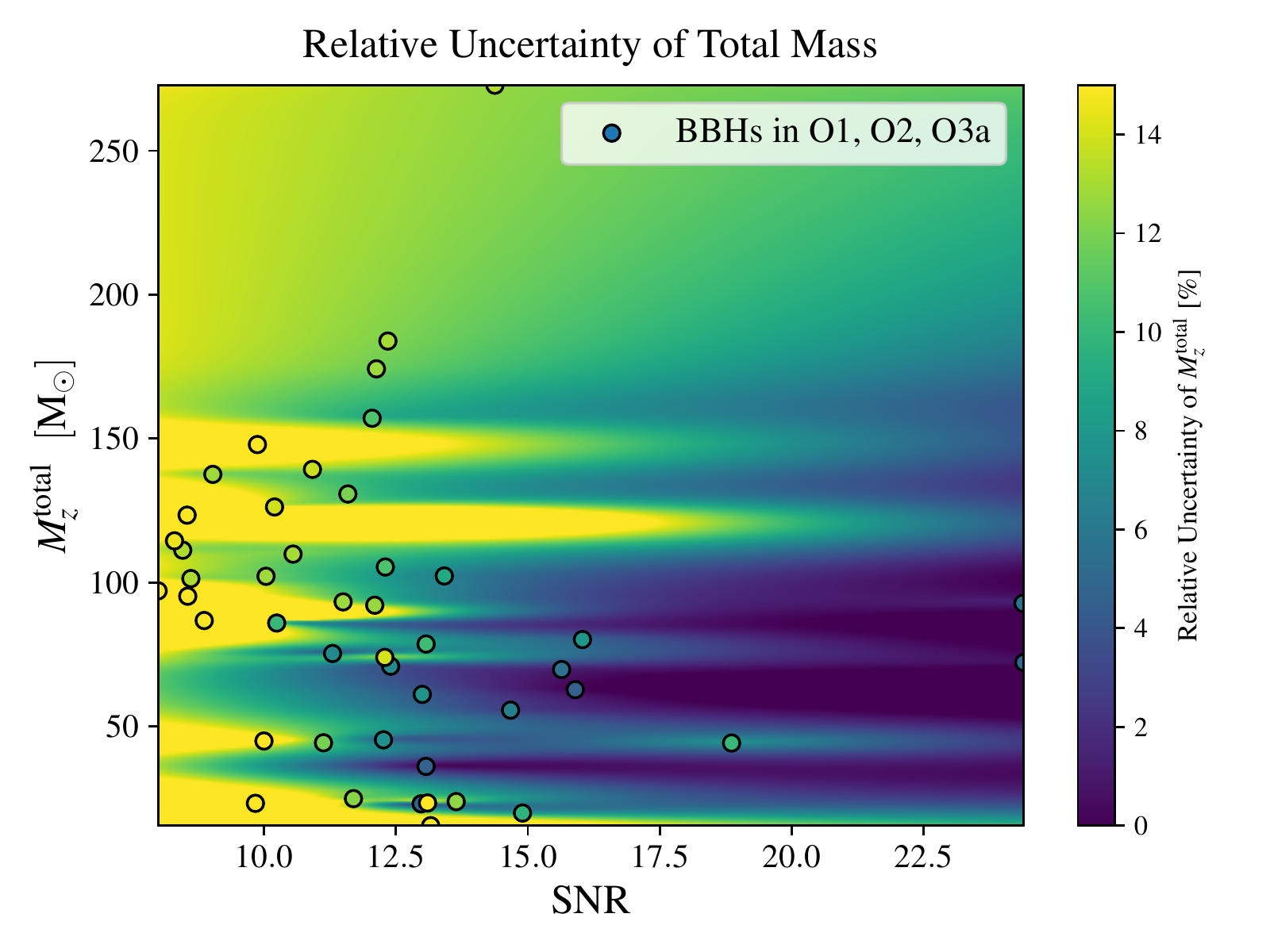}
    \caption{Contour map of relative uncertainty of detector-frame total mass as a function of detector-frame total mass and SNR. \textit{Horizontal-axis:} the SNR of the event. \textit{Vertical-axis:} the detector-frame total mass of the event. \textit{Right-bar:} relative uncertainty of the event (based on ($95\%$ confidence interval). Data points are BBHs in runs GWTC-2~\citep{Abbott_2019, Abbott_2021}. The color of the data points represents the actual relative uncertainty for the specific event. The rest of the contour plot is completed by interpolation. Any point with relative uncertainty higher than $15\%$ is also represented by the yellow color. Events with high masses generally have lower relative uncertainty than low mass events.
    }
    \label{fig:contour_total}
\end{figure}

To assess whether the chirp mass or the total mass performs better for low versus high mass events in terms of the false alarm probability, we analyze the relative uncertainty of chirp mass (total mass) as a function of chirp mass (total mass) and the SNR.
Figs.~\ref{fig:contour_chirp} and~\ref{fig:contour_total} show the contour maps for chirp mass and total mass, respectively. 
As seen in Fig.~\ref{fig:contour_chirp}, lower mass events generally have better-constrained chirp mass posteriors compared to higher mass events. Better-constrained posteriors will decrease the false alarm probability since the amount of coincidental overlap will be lower. Using chirp mass overlap instead of total mass overlap for lower mass events may be more beneficial in terms of the false alarm probability.

On the other hand, as seen in Fig.~\ref{fig:contour_total}, higher mass events generally have better-constrained total mass posteriors compared to lower mass events. Then, using the total mass overlap for higher mass events may be more beneficial in terms of the false alarm probability. In either case, it could be more beneficial to check the overlap of both of parameters (as opposed to only checking $\mathcal{M}_z$) when a lensing candidate is analyzed.

\subsection{Comparison of Uncertainties in Simulation and Catalog} \label{app:comparison_of_uncertainties}

\begin{figure}[htp]
    \centering
    \includegraphics[width=\columnwidth]{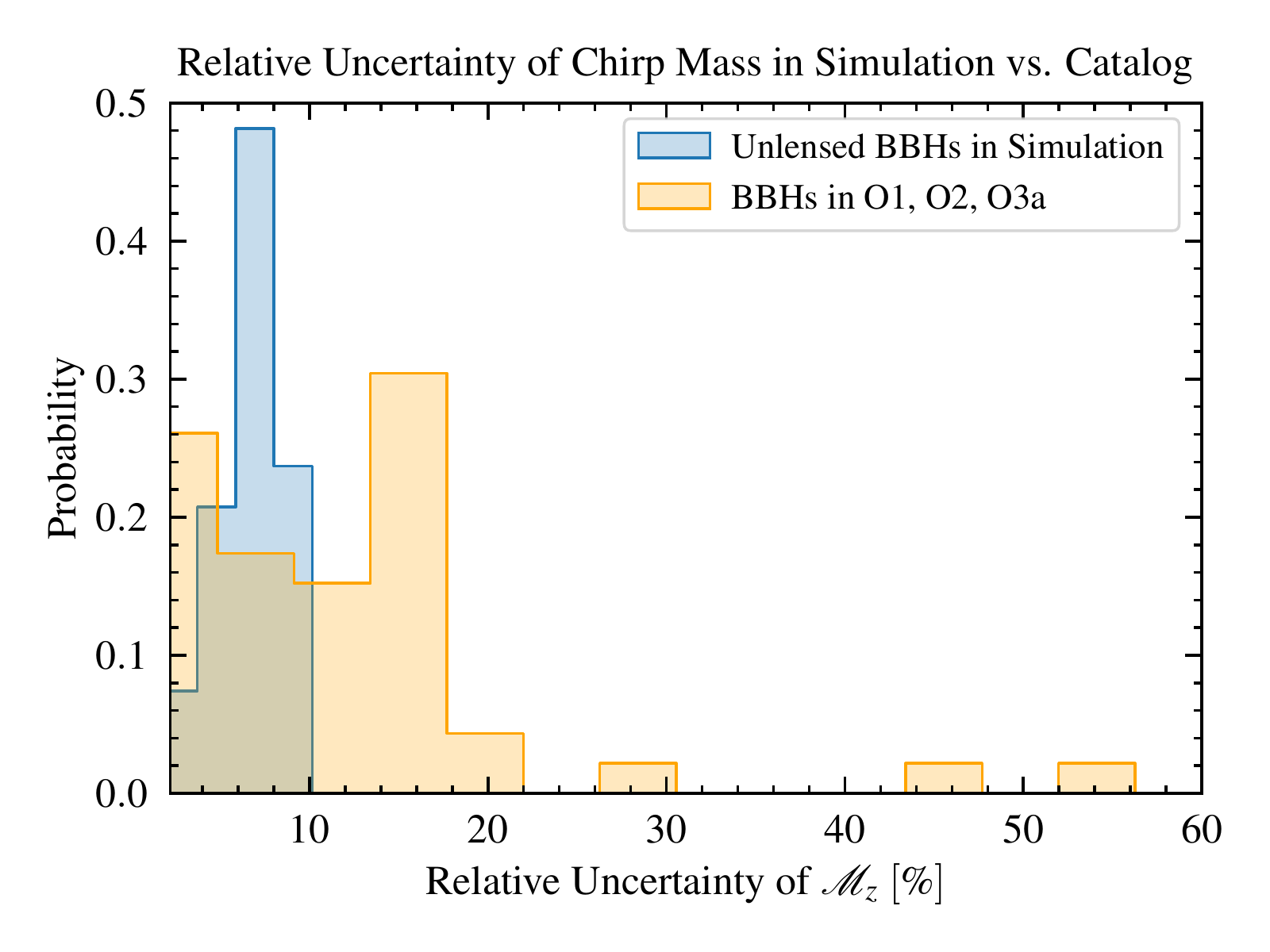}
    \caption{Probability density function of relative uncertainty ($95\%$ confidence interval) of the detector-frame chirp mass for the unlensed BBHs in simulation (blue) and the BBHs in GWTC-2~\citep{Abbott_2019, Abbott_2021} (orange).
    }
    \label{fig:uncertainty_comparison}
\end{figure}

Fig.~\ref{fig:uncertainty_comparison} shows the PDF of relative uncertainty (95\% confidence interval) of the detector-frame chirp mass for the unlensed BBHs in simulation and the BBHs in GWTC-2. We find that the distribution of the relative uncertainty of the simulated events is consistent with the events in the catalog. The catalog events have some high relative uncertainty events (e.g., $\sim45\%$) while the simulated events do not because we assume the simulated events are detected by the LIGO-Virgo detector network at design sensitivity.

\subsection{Comparison of SNRs in Simulation and Catalog} \label{app:comparison_of_SNR}

\begin{figure}[H]
    \centering
    \includegraphics[width=\columnwidth]{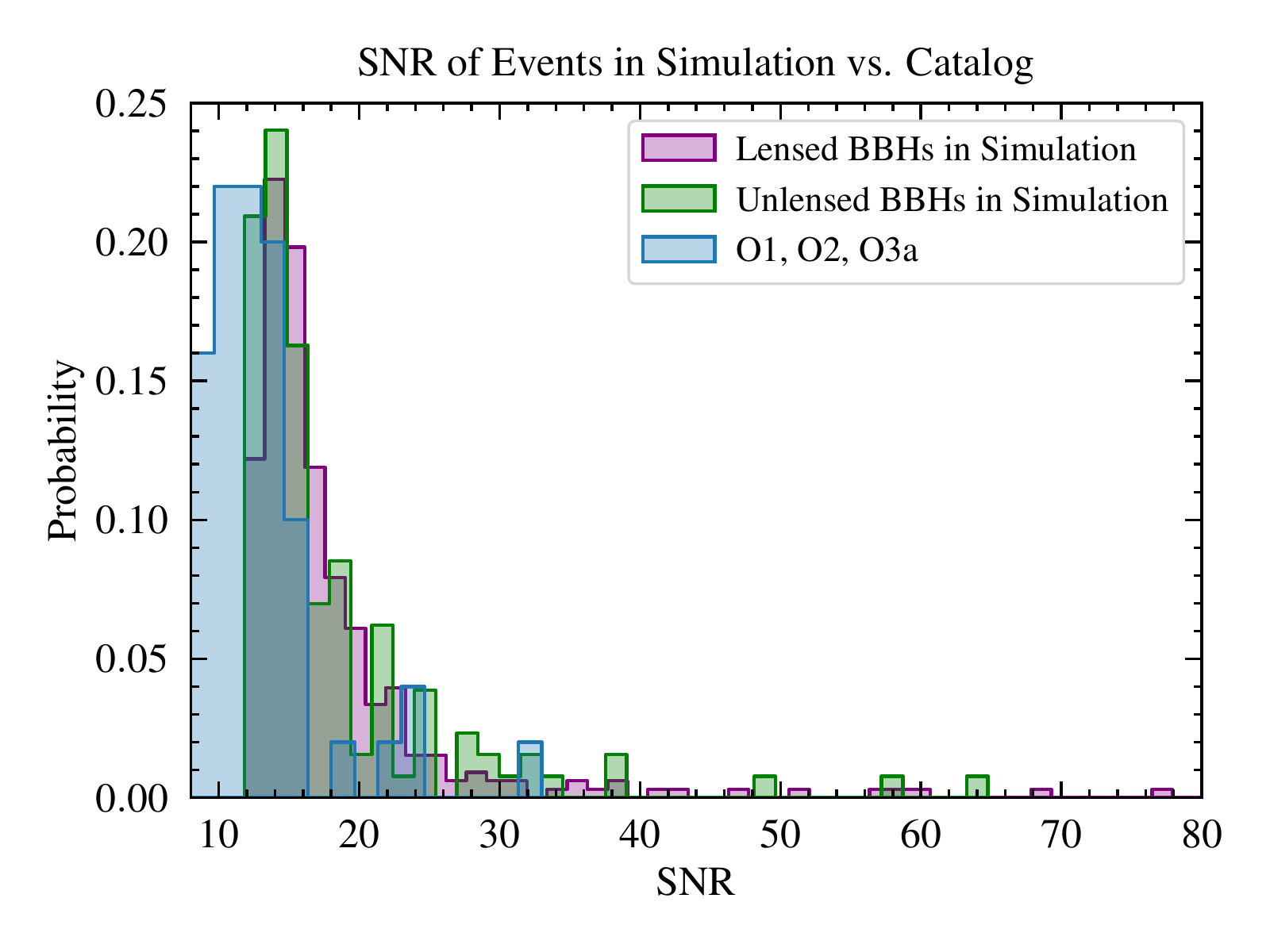}
    \caption{The probability function of the SNR of catalog events (blue)~\citep{Abbott_2019, Abbott_2021}, and lensed (purple) and unlensed (green) BBHs in simulation.
    }
    \label{fig:SNR_comparison}
\end{figure}

Fig.~\ref{fig:SNR_comparison} shows the PDF of the SNR for the unlensed and lensed BBHs in simulation and the BBHs in GWTC-2. Some of the simulated events have relatively higher SNRs. This is because we assume that the simulated events are detected by the LIGO-Virgo detector network at design sensitivity.

\section{GW measured phase} \label{app:measured_phase}
For a GW, the dominant (22) quadrupolar radiation has a form
\begin{equation}
    h_{22}=\mathcal{A}_{22}\cos(2(\Omega (t-t_{c})+\varphi_c)-\chi_{22})
\end{equation}
where 
\begin{equation}
    \chi_{22}=\rm arctan[{F_+(\theta,\phi,\psi)},
    f_{22}(\iota) {F_\times(\theta,\phi,\psi)}]\,,
\end{equation}
and 
\begin{equation}
    f_{22}(\iota) = \frac{2 \cos\iota}{1+\cos^2\iota}\,.
\end{equation}
The angle $\chi_{22}$ depends on the antenna pattern functions, which are given by 
\begin{align*}
    F_{+}&= \frac{1}{2}\left[1+\cos^2(\theta)\right]\cos (2\phi) \cos (2\psi) \\
    &-\cos(\theta)\sin (2\phi) \sin(2\psi),\\
    F_{\times} &= \frac{1}{2}\left[1+\cos^2(\theta)\right]\cos (2\phi) \sin (2\psi) \\
    &+\cos(\theta)\sin (2\phi) \cos(2\psi).
\end{align*}
Therefore, the particular phase that is well measured by LVK is 
\begin{equation}
    \Phi\equiv2\varphi_c - \chi_{22}(\theta,\phi,\psi,\iota)\,,
\end{equation}
rather than $\varphi_c$. 
To test the lensing hypothesis, we need to check if $\Delta \Phi = 0, \pi/2, \pi$ \cite{Ezquiaga:2020gdt}. 
One can use Fig.~\ref{fig:phase_overlap} to infer the \FAPpair for the phase overlap of $\Phi$ given an estimate of $\sigma$. Using a better-constrained phase will help reduce the \FAPpair coming from the phase.

%---------
%BIBLIOGRAPHY
%---------

%\bibliographystyle{unsrt}
\bibliography{lensingFAP}
\end{document}